\documentstyle[galley,epsfig]{mn2e}
\title{Subaru/HDS study of  CH stars: elemental abundances for  stellar neutron-capture process studies}
\title[Subaru/HDS study of   CH stars: elemental abundances for stellar neutron-capture process studies]
{Subaru/HDS study of  CH stars: elemental abundances for stellar  neutron-capture process studies\thanks{Based on data collected at the Subaru Telescope, which is
operated by the National Astronomical Observatory of Japan } }
\author[Aruna Goswami et al. ]{Aruna Goswami$^{1}$, Wako Aoki$^{2}$, 
Drisya Karinkuzhi$^{1}$
\\
$^{1}$Indian Institute of Astrophysics, Koramangala, Bangalore 560034, India;
aruna@iiap.res.in; drisya@iiap.res.in \\
$^{2}$  National Astronomical Observatory, Mitaka, Tokyo, 181-8588 Japan; 
aoki.wako@nao.ac.jp
}

\begin{document}
\date{  Accepted :  Received :  in original form :}

\pagerange{\pageref{firstpage}--\pageref{lastpage}} \pubyear{2011}

\maketitle

\label{firstpage}

\begin{abstract}
A comprehensive abundance analysis  providing  rare insight 
into the chemical history of lead stars is still lacking. We present results 
from high resolution ($R \sim 50\,000$), spectral analyses of three CH stars,  
HD~26, HD~198269, HD~224959, and, a carbon star with a dusty envelope,
HD~100764. Previous studies on these objects are  
limited by both resolution  and wavelength regions and the  results 
 differ significantly from each other. We have  undertaken 
to re-analyse the  chemical composition of these objects based on  high
resolution Subaru spectra covering the  wavelength regions 4020 to 6775 \AA\,.
Considering   local thermodynamic equilibrium and using model atmospheres, 
we have derived the stellar parameters, the effective temperatures 
T$_{eff}$, surface gravities log\,g, and metallicities ${\rm [Fe/H]}$ 
for these objects. The derived parameters for HD~26, HD~100764, HD~198269 
and HD~224959 are (5000, 1.6, $-$1.13), (4750, 2.0 $-$0.86), 
(4500, 1.5, $-$2.06) and  (5050, 2.1, $-$2.44) respectively. The stars 
are found to exhibit large enhancements of heavy elements relative to 
iron in conformity to previous studies. Large enhancement of Pb with 
respect to iron is also confirmed. Updates on the elemental abundances 
for several s-process elements (Y, Zr, La, Ce, Nd, Sm, Pb) along with 
the first-time estimates of abundances for a number of other heavy 
elements (Sr, Ba, Pr, Eu, Er, W) are reported. 
Our analysis suggests that  neutron-capture elements in HD~26 
primarily originate in s-process while the major contributions to the 
abundances of neutron-capture elements in the more metal-poor objects  
HD~224959 and HD~198269 are from r-process, possibly formed from materials
that are  pre-enriched with products of  r-process. 
\end{abstract}

\begin {keywords}
 stars: individual (HD~26, HD~198269, HD~224959, HD~100764)\,-\, 
stars: carbon\,-\,  stars: CH\,-\,  stars: CEMP-r/s\,-\,
  stars: Abundances\,-\,  stars: Nucleosynthesis
\end{keywords}

\section{Introduction}
Slow neutron-capture (s-process) elements are mainly produced by thermally 
pulsating asymptotic giant branch (AGB) stars; however, there remains 
large uncertainties associated with the formation of the main neutron 
source $^{13}$C. Observations of stars that show enhancement of 
s-process elements abundance can provide important constraints on 
theoretical models. CH stars (Keenan 1942) form good candidates to 
investigate the properties of s-process  at low-metallicities as their 
spectra exhibit  weak lines of the iron-group elements but enhanced 
lines of heavy elements. The stellar population of the Galactic halo 
is characterized by metal-poor low-mass stars, in which,  Carbon-Enhanced 
Metal-Poor (CEMP) stars form a substantial fraction of the objects 
with metallicity, [Fe/H] $\le$ $-$2. About 80\% of the CEMP stars
are known to be enriched with  s-process elements (Lucatello et al. 2003,
Aoki et al. 2007, Roederer et al. 2008). These objects are more metal-poor 
counterparts of CH stars. The  chemical composition  of these stars 
offer insights into the nucleosynthesis of heavy-elements. Chemical 
composition studies of CEMP stars (Barbuy et al. 2005, Norris et al. 
1997a, b, 2002; Aoki et al. 2001, 2002a, b; Goswami et al. 2006, 
Goswami \& Aoki 2010, Karinkuzhi \& Goswami 2014, 2015) have suggested that a 
variety of production mechanisms are needed to explain the observed range of 
elemental abundance patterns in them; however, the binary scenario of CH
star formation is currently considered as the most likely formation mechanism
 also for CEMP-s stars. Lucatello et al. (2005b) have demonstrated that 
all CEMP-s stars are likely members of binary systems. However, recent 
models of binary population synthesis are unable to reproduce the 
observed fraction of CEMP stars without invoking non-standard 
nucleosynthesis or a substantial change in IMF (Abate et al. 2013).
There exist different scenarios of carbon and s-process element production
at low and near-solar metallicity stars. For the origin of the s-process 
isotopes with mass number A $\le$ 90, the  activation of the 
$^{22}$Ne($\alpha$,n)$^{25}$Mg reaction appears the main production 
mechanism (Prantzos et al. 1987). This is referred as weak-s-process
mechanism.  For the production of the s-process nuclei with A $\ge$ 90 
(main component), the operation of the $^{13}$C($\alpha$,n)$^{16}$O in 
low-mass stars of rather low-metallicity seems to be a good site 
(Hollowell and Iben 1988).  Neutron-capture process characterized by 
high flux of neutrons (r-process) is hypothesized by Truran (1981) 
to be responsible for the production of most neutron-capture elements 
at the early ages of the Galactic life. 
For the  enhanced abundance of 
Eu with [Eu/Ba] ratios considerably higher than that resulting from 
s-process nucleosynthesis calculations, (these stars are classified 
as CEMP-r/s stars, Beers \& Christlieb 2005), no convincing explanation  
is yet available. 
Except for the 
overabundance of Eu, the CEMP-r/s stars do not show any systematic 
difference from the CEMP-s stars. Although considerable theoretical 
and observational interest has been focussed on these stars much 
remains mysterious about them  (i.e., Zijlstra (2004)). 
 Our understanding of the range of 
physical phenomena involved in the formation processes of these 
objects  still remains incomplete and needs to be substantiated by 
further observational and theoretical studies.
 In this work, we have  undertaken to conduct   fresh analysis of  
chemical  compositions for   stars HD~26, 198269 and 224959 listed
in CH star catalogue of Bartkevicious (1996) and HD 100764, a carbon star 
known to have a dusty envelope. Previous studies on these objects are 
limited by both resolution and spectral coverage  and  differ 
significantly  from each other. The fact that the results reported 
by Van Eck et al. (2003) for heavy elements such as Zr, La, Ce and Nd 
in  three of the four  CH stars  which are based on  better spectra are 
very different from  those reported by Vanture (1992c) motivated us 
to re-estimate the chemical abundances,   examine  the  extent to 
which   our results confirm or contrast the previous results and 
hence understand their chemical history. It is towards this end that 
we have undertaken to conduct  this study  based on high resolution 
spectra obtained using HDS attached to 8-m Subaru telescope  and  
re-evaluate the atmospheric parameters and the  chemical compositions.
The presence of several s-process elements  such as Sr, Y, Zr, Ba, La, 
Ce, Nd, Sm are clearly evident from  these spectra.  We report new 
results on the chemical composition  and discuss  mechanisms  giving 
rise to the overabundance of  s-process elements in light of  existing 
theories.

Models of very low-metallicity AGB stars developed   in the frame work  of
the partial mixing of protons into the deep carbon rich layers, predict 
overabundances of Pb-Bi as compared to lighter s elements  (Goriely \&
Mowlavi 2000). These stars,
are characterized by large [Pb/Fe], [s/Fe] and [Pb/s] abundance ratios, where
s represents s-process elements. In this scenario, AGB stars with 
[Fe/H] ${\le}$ $-$1.3, are predicted to exhibit [Pb/hs] (where hs, refers 
to heavy s-process elements, such as Ba, La, Ce) ratios as large as 1.5. 
These predictions are found to be quite robust with respect to the model 
parameters, i.e., the abundance profile of the protons in the partially 
mixed layers or the  extent of the partial mixing zone and uncertainties 
in  reaction rates. Low-metallicity AGB stars that exhibit large 
overabundances of Pb-Bi compared to lighter  s-elements are called as 
`lead-stars'. Although, HD~26, HD~198269 and HD~224959 are known as 
`lead-stars', the abundance of the  second-peak s-process element Ba, 
as well as other heavy elements such as Sr, Pr, Eu  and Er etc. are not 
available in literature for these objects. We have determined the 
abundances of these elements from analyses of their  high-resolution 
Subaru spectra. Radial velocity survey of McClure (1984), McClure and 
Woodsworth (1990), have confirmed binarity for three of  these  objects. 
As the chemical compositions  of these  stars can   provide observational
constraints for testing and building models of AGB  nucleosynthesis, 
accurate determination of the stellar parameters and their chemical 
compositions is  extremely important. Further, as these objects
occupy a wide range in metallicity, the question of metallicity 
dependence of s-process can also be investigated considering a 
large sample of these objects and covering a wide range in metallicity.

Recent studies have indicated that classical CH stars and CEMP-s stars,
probably belong to the same category (Masseron et al. 2010). If so, the 
population of these objects with accurate chemical history would be vital 
for understanding the contribution of low and intermediate mass stars 
to the chemical evolution of the Galaxy and other stellar systems.

In the next section we give a  brief summary of the previous studies
on these objects from literature and highlight the novel aspects of the
present study. In section 3, we present details of  observations, data 
analysis and estimates of radial velocities.  In  section 4, we  
present $BVRIJHK$  photometry of these stars, and give estimates of  
effective temperatures based on photometry.  The methodology followed  
for deriving  the stellar atmospheric parameters is discussed in section 5. 
The elemental abundance results are presented in section 6. Parametric 
model based analysis of the observed abundances is discussed in  section 7.   
Discussion  and concluding remarks are drawn in section 8. 

\section{Previous studies - a summary }
 Some  aspects of these objects were addressed by different authors 
in previous studies;   here we  summarize their main results.

{\bf  Lee (1974):} Lee carried out a detailed study on the chemical 
composition of  HD~198269 based on  three spectrograms having  dispersion
of 6.5 \AA\, mm$^{-1}$, 10 \AA\, mm$^{-1}$, and 178 \AA\, mm$^{-1}$
respectively,  obtained using KPNO 16 inch  telescope. As the available 
data was not sufficient to warrant a detailed model atmosphere approach, 
a standard curve-of-growth analysis was performed. The analysis was 
done with respect to  ${\epsilon}$~Vir, correcting  line strengths for the
difference in the effective temperatures of ${\epsilon}$~Vir and HD~198269.
 A metallicity ([Fe/H]) of  $-$1.56 was  adopted  for this object. At this 
metallicity [C/Fe] was estimated to be  +1.92.

{\bf Vanture (1992a, 1992b, 1992c):} In a series of three papers Vanture 
had discussed carbon isotopic ratios, abundances of C, N, O and heavy 
elements (also  Cu and Mo for HD~26) derived from an LTE analysis of 
the equivalent widths of weak atomic lines using modified version of 
MOOG (Sneden 1973). The analyses  were based on spectra observed with  
the 1.5-m telescope at the Cerro Tololo Inter American Observatory using 
the Fibre Fed Echelle with GEC CCD detector having  resolutions of 
0.24 \AA\, in the blue and 0.44 \AA\, in the red 
($\lambda/ \delta \lambda$ $\sim$ 20,000) with a S/N $\sim$ 80 - 150 in 
both red and  blue. Additional spectra for HD~26 and HD~224959 were  
obtained from 4-m telescope at Kitt Peak Observatory with  resolution of 
0.25 \AA\, at 8000 \AA\, ($\lambda/ \delta \lambda$ $\sim$ 32,000) and 
 S/N in the range 100 - 180.

Carbon isotopic ratios ($^{12}$C/$^{13}$C) were determined from 
spectrum synthesis calculations  using the 1-0 C$_{2}$ Swan system 
at 4737 \AA\, and the CN red system 2-0 band near 8000 \AA\,.
The derived values are $>$ 25 for  HD~26 (from both C$_{2}$ and CN) 
which is about a factor of 2 higher than that of Aoki \& Tsuji (1997).  
For HD~198269, these estimates are respectively 4 and 6 and for 
HD~224959, 3 and 13. 

Vanture used  the Oxygen triplet lines of 7771, 7774 and 7775 \AA\, 
to determine the abundances of Oxygen.  These lines returned  abundances 
that differed by 0.3 dex to 0.6 dex. Weak undetected lines blending 
with the oxygen lines was cited as  the primary reason causing this 
difference. Spectrum synthesis method was also  used around 6300 \AA\, 
region to constrain  the oxygen abundance in HD~198269 and HD~224959. 
As the spectrum of HD~26 around [O I] 6300 \AA\, was not of high 
quality to set an upper limit for oxygen, the oxygen abundance for 
this star was adopted from near IR oxygen triplet lines alone.

Using oxygen abundance derived from O I triplet lines, carbon abundance in
HD~26 was derived from spectral synthesis calculation of C$_{2}$  molecular
band at 4737 \AA\,. For the other two stars carbon abundances were 
determined from analyses of  equivalent widths of the 
CH A$^{2}$${\Delta}$ - X$^{2}$${\Pi}$${\Delta}\nu$= $-1$1 lines at 4835 
and 4880 \AA\,. Carbon abundance derived from CH lines were further 
checked by synthesizing the C$_{2}$(1,0) band at 4737 \AA\,. Nitrogen 
abundance was  determined from spectral synthesis of the CN Red system 
in the regions 7995 - 8013 \AA\, and 8029 - 8042 \AA\,. Final N abundance 
adopted is the average of the two values derived from these two spectral 
regions.

C, N, O abundances (log ${\epsilon}$) for HD~26, derived with adopted
metallicity ([Fe/H]) of  $-$0.44 are respectively 8.4, 7.9 and 8.45. 
These abundances  are 8.5, 8.2 and $<$ 7.7 for HD~224959 and 7.9, 7.6 
and $<$ 7.6 respectively for HD~198262; the  adopted [Fe/H] for 
HD~224959 and HD~198269 were respectively $-$1.6 and $-1.4$. We note 
that, Vanture's  adopted  metallicities  vary  significantly from our 
estimates of $-$1.13,  $-$2.04 and $-$2.42 respectively for HD~26, 
HD~198269 and  HD~224959.

{\bf Aoki \& Tsuji (1997):} These authors have derived  carbon isotopic 
ratios ($^{12}$C/$^{13}$C) from spectral analysis of CN red system 
for HD~224959, HD~198269, and HD~100764. The values are respectively 
7, 7 and 4. This ratio estimated using C$_{2}$ Swan system is $\sim$ 10 
for HD~26. These values are  estimated at effective temperatures
that compare closely to our estimates. However, metallicities differ 
significantly. Low values  ($<$ 10) of ($^{12}$C/$^{13}$C) qualify the 
objects to be early type CH stars that are post-mass-transfer binary 
systems (McClure \& Woodsworth 1990).  In Table 1, we have summarized
the  literature values of C, N, O abundances with respect to Fe as well
as $^{12}$C/$^{13}$C ratios.

{\bf Van Eck et al. (2001, 2003):}  The first detection of lead 
in HD~26, HD~198269 and HD~224959 were reported in these papers.
 Abundances for C, heavy elements Zr, La, Ce, Nd and Sm were also  
presented.  The results are based  on two separate observations; 
i)  medium resolution spectra of two spectral windows containing 
Pb lines obtained at  Observatoire de Haute Provence (OHP) in 
Aug 2000; ii) high-resolution spectra  obtained at ESO in  Sep 2000.  
The OHP spectra were obtained with 1.52m telescope at a resolution 
of $\lambda/ \delta \lambda$ $\sim$ 85,000 in the 2nd order at 720 nm 
to explore the 722.897 nm Pb I line and 3000 rule mm$^{-1}$ holographic 
grating was  used to explore the 401.963, 405.781 and 406.214 nm Pb I 
lines with a resolution of 45000. The ESO spectra were obtained on 
the Coude Echelle Spectrometer (CES) fed by the 3.6-m telescope. 
A resolution of 135,000 was achieved at the central wavelength of 
405.8 nm. The spectra covered a wavelength range  of 404.5 - 407.1 nm.

The s-scatter rather than the r-process was believed to be responsible 
for the observed overabundances of heavy elements.  The authors 
concluded that  HD~224959 and HD~198269 belong to the class of 
`lead stars', since their [Pb/Fe] ratios comply with the prediction
for the standard model for Partial Mixing of Protons (PMP) operating in 
low-metallicity AGB stars.
\begin{table*}
{\bf Table 1: CNO abundances from the literature} \\

\begin{tabular}{|l|c|c|c|c|c|c|c}
\hline
Star& $^{12}C/^{13}C$ & $^{12}C/^{13}C$  &[C/Fe]& [N/Fe]  & [O/Fe]& [O/Fe]&Ref\\
    &  (4737 \AA)   &    (8000 \AA)&               &          &  (6300 \AA)   & (7777 \AA)     &    \\
\hline
HD~26     &-     & -    & 0.31 & 1.12&0.34& &1\\ 
     &$\ge$ 25     & $\ge$ 25    & 0.41 & 0.51&-& 0.20&2\\ 
    &-     &9   & 0.68  &0.94 &0.36&- &3\\
HD~198269&    -     &  -   &  2.31 &1.77&0.33&- &1\\
&     -    &   -  &  1.92 &- &-& -&4\\
&   4      &  6   &  0.87 & 0.91&-&0.05 &2\\
HD~224959&-     &  - &   2.01 & 1.98& 0.34&- &1\\
&3     &  13&  1.67 & 1.97&- &0.61&2\\
&-   &  4  &  1.77 & 1.88&1.10& -&3\\
\hline 
\end{tabular}

1.This work, 2. Vanture (1992a), 3. Masseron et al. (2010), 4. Lee (1974) \\
\end{table*}

{\bf Novel aspects of the present work:} The present results are based on 
a detailed abundance analysis  based on high-resolution (R $\sim$ 50,000) 
spectra obtained with Subaru using HDS (Noguchi et al. 2002). The wavelength
coverage is continuous from 4020 through 6775 \AA\, except for a small 
gap of about 75 \AA\, from 5335 \AA\, to 5410 \AA\,.   Van Eck et al. 
used a higher resolution spectra for obtaining Pb abundances but the
wavelength coverage is limited to 4045 - 4071 \AA\,. In determining the 
stellar atmospheric parameters (T$_{eff}$, log g, and [Fe/H]), from 
an LTE analysis  of the equivalent widths of atomic lines of Fe, we 
have used the most recent version of the spectrum synthesis code MOOG  
(Sneden 1973) with standard assumptions of  Local Thermodynamic Equilibrium,
 plane parallel atmosphere, hydrostatic equilibrium and flux conservation.
We have made use of the new grid of ATLAS09 model atmospheres of Kurucz
database (http://kurucz.harvard.edu/grids.html).  Our analysis is also 
benefited by the fact that we have used an extensive line-list for heavy
elements  (Table 13) in consultation with the most updated and
well tested log gf values.  We have measured the equivalent widths for 
a large number of lines in the solar spectra, that are common in our program 
stars spectra and measured
the solar abundances using the  log gf values that we have adopted.  The
solar atmospheric parameters used are T $_{eff}$ = 5835 K, 
log g = 4.55 cm s$^{-2}$  and micro turbulent velocity 1.25 km s$^{-1}$. 
The derived
abundances match closely within a range of 0.08 - 0.1 dex, when comparing with
Asplund (2009). 
We have presented the first-time estimates
of abundances for the heavy elements Sr, Ba, Pr, Eu, Er and W.
We have also performed a parametric model based analysis of the 
observed abundances to  examine the relative  contributions from  
the s- and r-process to  the observed abundances of heavy elements 
in order to understand the underlying nucleosynthesis responsible for 
the observed abundances.

\section{Observation, Data Reduction, and Radial Velocities}

High-resolution spectra for the objects were obtained using the High 
Dispersion Spectrograph (HDS) attached to the 8.2m Subaru
Telescope (Noguchi et al. 2002) on  May 30, 2004 (HD~224959) and
June 1, 2004 (HD~26, HD~100764 and HD~198269). Each object spectrum was
taken with a 5 minutes exposure having a resolving power of $R \sim$
50\,000. The observed bandpass ran from about $4020$\,{\AA} to
$6775$\,{\AA}, with a gap of about $75$\,{\AA}, from $5335$\,{\AA} to
$5410$\,{\AA}, due to the physical spacing of the CCD detectors. The data
 reduction was  carried out, in the standard fashion, using
IRAF\footnote{IRAF is distributed by the National
Optical Astronomical Observatories, which is operated by the Association
for Universities for Research in Astronomy, Inc., under contract to the
National Science Foundation} spectroscopic reduction package.
The basic parameters of these objects are listed in Table 2.

{\footnotesize
\begin{table*}
{\bf Table 2: Photometric parameters for program and comparison stars}\\
\tiny
\begin{tabular}{ l c c c c c c c c c c c c }
  &   &   &  &  &   &  &  &   &   &  \\
\hline
Stars   &RA(2000)  & Dec(2000) & Sp Ty  &  V   & $B-V$   &$R-I$   & $V-I$   &$E(B-V)$
&  $J$    &$ H$  & $K_{s}$\\
  &   &   &  &  &   &  &  &   &   &  \\
\hline
HD~26     &00~05~22.20 &+08~47~16.11 &K0III/G4Vp & 8.22 & 1.05 & -    & -  & 0.05 & 6.540 & 6.106 & 6.032 \\
HD100764  &11~35~42.74 &-14~35~36.66 &CH/R0/R2/C & 8.73 & 1.05 & 0.51 & 0.92 & 0.02 & 7.048 & 6.60 & 6.513 \\
HD198269  &20~48~36.74 &+17~50~23.72 &R0/R       & 8.12 & 1.28 & 0.58 & 1.14 & 0.10 & 6.078 & 5.505 & 5.385 \\ 
HD224959  &00~02~08.02 &-02~49~12.26 &R2/R0      & 9.55 & 1.12 & 0.51 & 0.96 & 0.03 & 7.863 & 7.432 & 7.303  \\
  &   &   &  &  &   &  &  &   &   &  \\
\hline
\end{tabular}

The values of B-V, R-I, V-I are taken from Platais et al. 2003, A\&A, 397, 997.
\end{table*}
}
{\footnotesize
\begin{table*}
{\bf Table 3:  Heliocentric Radial velocities  v$_{r}$  of the program stars}\\
\begin{tabular}{ l c  c  c  c  }
\hline
  &   &   &  &    \\
Stars     &      HJD        & $v_{\rm r}$ km s$^{-1}$ &  $v_{\rm r}$ km s$^{-1}$ \\
          &                 &  Our estimates          & from literature  & References   \\ 
  &   &   &  &    \\
\hline
  &   &   &  &    \\
HD~26     &   2452784.95121 & +210.5 ${\pm}$ 1.5      & $-$212.9   &  1  \\
HD~100764 &   2453157.73074 &  +4.9 ${\pm}$ 1.5       &    +5.0    & 2 \\
HD~198269 &  2453158.01437  & $-$200.5 ${\pm}$ 1.5    & $-$204.0 &  3  \\
          &                      &                    & $-$207.0 & 4 \\
          &                      &                    & $-$203.4 & 5 \\
HD~224959 &  2453156.12808  & $-$126.5 ${\pm}$ 1.5    & $-$132.0   & 4 \\
          &                 &                         & $-$127.8   & 5 \\
\hline
\end{tabular}

References-  1 : Aoki \& Tsuji (1997), 2: Keenan (1993) and Dominy (1984),
  3 : Nordstroem, B. et al. (2004),
4: Hartwick \& Cowley (1985), 5: McClure \& Woodsworth (1990) \\
 
\end{table*}
}

\subsection{Radial velocities}

The radial velocities of our program stars are  measured using several
unblended lines.  Estimated heliocentric radial velocities $v_{\rm r}$ of
the program stars are presented in Table 3. Previous estimates of 
radial velocities for these objects available in literature are also 
presented for a comparison.  Except for HD~100764 with a radial 
velocity of about  5 km s$^{-1}$, the other three are high velocity
objects. The heliocentric radial velocities for HD~26, HD~198269 and 
HD~224959 are respectively $+$210 km s$^{-1}$, $-$200 km s$^{-1}$ and 
$-$126 km s$^{-1}$.

\section{Results}

The high-resolution spectra  are characterized by  closely-spaced  
lines of carbon bearing molecules of CH, CN and the Swan system 
of C$_{2}$.  A few unblended atomic lines were possible to identify 
only from the regions that are relatively clear of molecular lines. On 
a  line-by-line basis, possible  blends between atomic and molecular 
lines  were identified and eliminated and  unblended atomic lines were 
measured from the regions that are relatively clear of molecular lines.
 Both the CH  and the C$_{2}$ Swan band systems 
line lists   of Phillips and Davis (1968), the solar atlas,  and,  
 the atomic line lists of Kurucz were used 
for this purpose.  A  list of Fe I and Fe II lines considered in the 
present analysis  for the determination of atmospheric
parameters and metallicity  is given in Table 4.

{\footnotesize
\begin{table*}
{\bf Table 4:  Fe lines used for determination of  the atmospheric parameters}\\
%\tiny
\begin{tabular}{ |c|c|c|c|c|c|c|c| }
\hline
         &    &           &            & HD~100764   & HD~224959    & HD~198269 & HD~26\\
W$_{lab}$& ID & EP$_{low}$& log gf  & Eqw (m\AA\,)& Eqw (m\AA\,) & Eqw (m\AA\,)& Eqw (m\AA\,) \\
         &    &           &            &             &              &             &     \\
\hline
6593.871 & Fe I &  2.43 & $-$2.42    & ---   &  ---  &  75.7 & --- \\
6592.910 & Fe I &  2.72 & $-$1.60    & 140.5 &  ---  &  ---  & --- \\
6518.380 & Fe I &  2.83 & $-$2.75    &  78.5 &  ---  &  ---  & --- \\
6421.350 & Fe I &  2.28 & $-$2.03    &  ---  &  ---  &  111.3& --- \\
6411.650 & Fe I &  3.65 & $-$0.82    &  ---  & 34.9  &  88.9 & --- \\
6393.612 & Fe I &  2.43 & $-$1.62    &  ---  & ---   &  135.7& --- \\
6335.340 & Fe I &  2.19 & $-$2.23    & 134.0 & ---   &  ---  & ---\\
6252.550 & Fe I &  2.40 & $-$1.69    & 154.3 & 41.7  & ---   & ---\\
6246.317 & Fe I &  3.60 & $-$0.96    &  ---  &  ---  &  78.6 & ---\\
6230.726 & Fe I &  2.55 & $-$1.28    & 190.4 & 57.6  & ---   & ---\\
6219.279 & Fe I &  2.20 & $-$2.43    &   --- & ---   & ---   & ---\\
6137.694 & Fe I &  2.58 & $-$1.40    & 189.0 & ---   &  97.8 & ---\\
6136.615 & Fe I &  2.45 & $-$1.40    & 184.0 & 53.8  & 124.6 & --- \\
5586.760 & Fe I &  3.36 & $-$0.21    & 159.3 & 58.6  & 118.0 & --- \\
5324.179 & Fe I &  3.21  & $-$0.10   & ---   & 69.3  & 130.6 & 152.2\\
5266.555 & Fe I &  2.99 & $-$0.49    & ---   & ---   & ---   &  --- \\
5242.490 & Fe I &  3.63 & $-$0.84    &  98.6 & ---   & ---   & --- \\
5232.939 & Fe I &  2.94 & $-$0.19    &  ---  & 82.6  & 148.3 & 169.0\\
5202.340 & Fe I &  2.18 & $-$1.84    &  ---  & 47.7  & 129.7 & ---  \\
5192.343 & Fe I &  2.99 & $-$0.52    & ---   & 64.8  & 130.7 & ---\\
5171.595 & Fe I &  1.48 & $-$1.79    &  ---  & 86.9  & 156.8 & 157.6\\
5166.282 & Fe I &  0.00 & $-$4.19    &  ---  & 51.3  & 150.8 & 145.2\\
5051.635 & Fe I &  0.91 & $-$2.79    &  ---  & 84.0  & 157.2 & ---  \\
5049.819 & Fe I &  2.27 & $-$1.42    &  ---  &  ---  & 139.0 & ---  \\
5012.068 & Fe I &  0.85 & $-$2.64    & ---   & 74.8  & 169.3 & --- \\
5006.119 & Fe I &  2.83 & $-$0.61    &  ---  & 71.0  & 130.4 & --- \\
5001.860 & Fe I &  3.88 & $+$0.009   & 146.4 & 32.4  & ---   & 100.8\\
4994.130 & Fe I &  0.91 & $-$2.96    &   --- &  59.7 & 130.4 & 129.8\\
4982.524 & Fe I &  4.10 & $+$0.14    & 129.1 & ----  &  75.8 & --- \\
4966.087 & Fe I &  3.33 & $-$0.89    &   --- &   --- &  85.6 & 103.2\\
4939.690 & Fe I &  0.85  & $-$3.34   &  ---  &  ---  &  ---  & 130.0\\
4938.814 & Fe I &  2.87  & $-$1.08   &  ---  &  ---  &  ---  & 100.7\\
4924.770 & Fe I &  2.27 & $-$2.25    & 136.3 & ---   &  93.8 &  99.2\\
4918.990 & Fe I &  2.87 & $-$0.34    &   --- &  81.6 & 139.1 &  148.7\\
4903.310 & Fe I &  2.88 & $-$0.93    & 151.0 & 46.3  & 104.6 & 122.7\\
4891.490 & Fe I &  2.85 & $-$0.14    &  ---  & ---   & 151.5 & 179.2 \\
4890.755 & Fe I &  2.87  & $-$0.39   &  ---  & ---   & ---   & 181.5 \\
4877.610 & Fe I &  2.99 & $-$3.15    & 50.1  & ---   &  ---  & --- \\
4871.317 & Fe I &  2.86 & $-$0.41    &  ---  & ---   & 146.4 & 148.1 \\
4869.450 & Fe I &  3.54 & $-$2.52    & 49.2  & ---   & ---   & --- \\
4834.510 & Fe I &  2.42   & $-$3.41    & 70.0  & ---   & ---   &  --- \\
4787.833 & Fe I &  3.00   & $-$2.77    & 60.5  & ---   & ---   &  27.0 \\
4632.911 & Fe I &  1.61   & $-$2.91    &  ---  &  ---  &  96.5 &  --- \\
4494.560 & Fe I &  2.19   & $-$1.14    &  ---  &  70.7 & 130.0 & --- \\
4491.490 & Fe I &  2.85   & $-$0.11    &  ---  & --- &  ---  & --- \\
4489.740 & Fe I &  0.12   & $-$3.97    &  ---  &  ---  &  ---  & 134.7\\
4484.219 & Fe I &  3.60   & $-$0.72    & ---   & ---   & 67.6  &  88.5\\
4476.019 & Fe I &  2.84   & $-$0.82    &  ---  &  63.5 & ---   & 146.5\\
4466.552 & Fe I &  2.83   & $-$0.60    &  ---  &  ---  & ---   & 151.6\\
%         &      &         &              &       &       &       &   \\
\hline
\end{tabular}
\end{table*}
}

{\footnotesize
\begin{table*}
{\bf Table 4:  (continued)   }\\
\begin{tabular}{| c| c| c| c| c| c| c| c| }
\hline
         &    &           &            & HD~100764  &HD~224959    & HD~198269 & HD~26\\
W$_{lab}$& ID & EP$_{low}$ & log gf  & Eqw (m\AA\,)& Eqw (m\AA\,) & Eqw (m\AA\,)& Eqw (m\AA\,) \\
         &    &           &               &      &            &            &      \\
\hline
4447.720 & Fe I &  2.22   & $-$1.34    &  ---  &  ---  & ---   & --- \\
4443.190  & Fe I &  2.86   & $-$1.04    &  ---  &  50.0 & ---   & --- \\
4256.790 & Fe I &  4.25   & $-$1.56    & 45.1  & ---   & ---   & --- \\
6456.390 & Fe II&  3.90   & $-$2.07    &  ---  &   --- & 43.2  & --- \\
5234.625 & Fe II&  3.22   & $-$2.05    & 111.4 & 37.5  & 73.9  &  99.6 \\
5197.559 & Fe II&  3.23   & $-$2.10    & 106.6 & 48.5  & 76.8  &  --- \\
4993.350  & Fe II&  2.81   & $-$3.67    & ---   & ---   & ---   &  51.5\\
4923.930 & Fe II&  2.89   & $-$1.32    &   --- & 85.8  & 135.6 &  167.8\\
4731.440  & Fe II&  2.89   & $-$3.36    & ---   & ---  &  ---  & --- \\
4620.510 & Fe II&   2.83   & $-$3.29    &  ---  & ---   & ---   &  84.4\\
4583.839 & Fe II&  2.80   & $-$2.02    &  ---  & 72.6  & 106.5 &  --- \\
4576.330 & Fe II&  2.84   & $-$3.04    & 92.1  & ---   & ---   & ---  \\
4520.230  & Fe II&  2.81   & $-$2.60    & ---   &  58.4 & ---   & --- \\
4515.340  & Fe II&  2.84   & $-$2.48    & ---   &  29.7 & 67.9  & --- \\
4508.280 & Fe II&  2.85   & $-$2.21    & ---   &  44.3 & 78.6  & --- \\
4491.400 & Fe II&  2.85   & $-$2.70    & ---   &  27.9 & 59.2  & --- \\
\hline
\end{tabular}
\end{table*}
}

{\footnotesize
\begin{table*}
{\bf Table 5:  Effective temperatures from Photometry   }\\
\begin{tabular}{cccccclll}
\hline
Star Name&  V    &   J    &  H    &   K    & $T_{eff}$ (K) & $T_{eff}$ (K) &  $T_{eff}$ (K) & $T_{eff}$ (K) \\
         &       &        &       &        &  (J-K)      & (J-H)     &   (V-K)   &  Spectroscopy\\
\hline
HD 26   & 8.22 &  6.540 & 6.106 &  6.032  & 5112.4& 4957.8($-$0.5) & 4811.3($-$0.5) &   5000\\ 
&&&&&&4974.5($-$1.0)  &  4798.7($-$1.0)  &    \\
&&&&&&4991.3($-$1.5)  &  4790.7($-$1.5)   &    \\ 
&&&&&&5008.3($-$2.0)  &  4787.3($-$2.0)    &   \\ 
&&&&&&5025.3($-$2.5) &  4788.5($-$2.5)     &   \\
HD 100764 & 8.73 &  7.048 & 6.600 &  6.153 & 3849.9& 4990.9($-$0.5) & 4394.8($-$0.5) & 4750 \\
&&&&&&5007.6($-$1.0)  &  4378.6($-$1.0) &   \\
&&&&&&5024.4($-$1.5)  &  4366.2($-$1.5)  & \\
&&&&&&5041.3($-$2.0)  &  4357.6($-$2.0)  &  \\
&&&&&&5058.3($-$2.5) &  4352.8($-$2.5)   &  \\
HD 198269 & 8.12  & 6.708 & 5.505 & 5.385 & 2979.5& 2610.1($-$0.5) &  4245.6($-$0.5) & 4500  \\
&&&&&&2622.7($-$1.0)  &  4228.3($-$1.0) &    \\
&&&&&& 2635.3($-$1.5)  &  4214.6($-$1.5) &  \\ 
&&&&&&2648.1($-$2.0)  &  4204.5($-$2.0)   &  \\
&&&&&&2660.9($-$2.5) &  4197.9($-$2.5)   &   \\
HD 224959 & 9.55&   7.863 & 7.432 &  7.303 & 4902.6& 4988.6($-$0.5)&  4744.1($-$0.5 & 5000  \\
&&&&&&5005.3($-1$.0)  &  4730.8($-1$.0) &    \\
&&&&&&5022.1($-$1.5)  &  4722.1($-$1.5) &   \\
&&&&&&5038.9($-$2.0)  &  4717.7($-$2.0) &   \\ 
&&&&&&5056.0($-$2.5) &  4717.8($-$2.5)  &   \\
\hline 
\end{tabular}

The numbers in the parenthesis indicate  metallicities.\\
\end{table*}

\subsection{Temperature estimates from photometry}
Estimates of the effective temperatures of the  program stars have
been determined using the temperature calibrations derived by Alonso et al.
(1996) which relate T$_{\rm eff}$ with various optical and near-IR colours.
Using this method Alonso et al. estimate an external uncertainly of $\sim 90$K
in the temperature calculation.
The  calibrations of $B-V$, $V-R$, $V-I$, $R-I$ and $V-K$ require colours
in the Johnson system, and the calibrations of the IR colours $J-H$, and $J-K$
in the TCS system (the photometric system at the 1.54m Carlos Sanchez telescope
in Tenerife; Arribas \& Martinez-Roger 1987). To obtain the $V-K$ colour in the
Johnson system we first transformed the K$_{s}$ 2MASS magnitude to the TCS
system. Then, using eqs (6) and (7) of Alonso et al. (1994), K is transformed to
the Johnson system from the TCS system. In order to transform the 2MASS colours
$J-H$ and $J-K_S$ onto the TCS system we first transformed the 2MASS colours to
CIT colours (Cutri et al. 2003), and then from CIT to the TCS system (Alonso et
al. 1994).
Estimation of the T$_{\rm eff}$ from the T$_{\rm eff}$ - $(J-H)$ and T$_{\rm
eff}$ - $(V-K)$ relations also involves a metallicity ([Fe/H]) term. We have
estimated the $T_{\rm eff}$ of the stars at several metallicities. The estimated
temperatures, along with the adopted metallicities, are listed in Table 5.
The temperatures  span a wide range. In particular,
in the case of HD~198269, $J-H$ and $J-K_S$ provide very low temperatures than
that given by   $V-K$ colour.

\section{Stellar atmospheric parameters}
A standard LTE analyses procedure was adopted to determine the 
 stellar atmospheric parameters, the effective temperature
($T_{\rm eff}$), the surface gravity (log g), and metallicity
([Fe/H]) of the stars.  A  recent version of MOOG of Sneden
(1973) is used.    Model atmospheres are
selected from the Kurucz grid of model atmospheres computed with
better opacities and abundances  with no convective overshooting.
 These models are available at {\tt http://cfaku5.cfa.harvard.edu/},
labelled with the suffix ``odfnew''.  The abundance analysis is made
using the grid calculated for C-normal chemical composition.
   The excitation potentials and
oscillator strengths of the lines are  taken from various sources
 (Vienna Atomic Line
Database ({\tt http://ams.astro.univie.ac.at/vald/}),  Kurucz atomic
line list ({\tt
http://www.cfa.harvard.edu/\-amp/\-ampdata/\-kurucz23/\-sekur.html}),
Fuhr, Martin, \& Wiese (1988), Martin, Fuhr, \& Wiese (1988), and Lambert
et al. 1996,  $gf$ values of  elements compiled by
 R.E.  Luck.

The effective temperature of the stars were  obtained by the method of
excitation balance, forcing the slope of the abundances from Fe~I lines
versus excitation potential to be near zero. The temperature estimates
derived from JHK photometry (Table 5)  provided a preliminary temperature
 check for choosing an initial model atmosphere. The final effective 
temperatures were then obtained by an iterative process using the 
method of excitation balance. A comparison of the effective temperatures 
derived from photometry and spectroscopy shows that the spectroscopic 
estimates are closer to temperature estimates obtained from (J-H) 
relation which are not too different from those obtained from (J-K) 
relation.  HD~198269 is an exception where estimates from (J-K) and 
(J-H) show significant difference from the spectroscopic value. It is 
likely that in case of HD~198269 there could be some molecular bands 
affecting J band. The T$_{eff}$ from (V-K) show a deviation
of about 300 K from spectroscopic ones, the estimates from (V-K) is
relatively reliable compared to those  from optical colours e.g.
V-R and B-V.  The adopted microturbulence is 2 km s$^{-1}$ for all the 
objects. Such a value  is not unrealistic; (in cool giants, with 
log\,g $\le$ 2.0, in general $V_{\rm t}$  ${\ge}$ 2 km s$^{-1}$  
(Vanture 1992c, McWilliam et al. 1995a,b)).  The surface gravity of the 
stars were derived using  the Fe~I/Fe~II ionisation equilibrium. The 
derived atmospheric parameters $T_{\rm eff}$, log\,g, and [Fe/H] of 
the stars are listed in Table 6. The correctness of our estimates is 
verified by reproducing the atmospheric parameters obtained by 
Barbuy et al. (2005)  for the star CS~22948-027. Their  estimates are  
very close to our  estimates  (refer to Goswami et al. (2006), Table 5).  

\section{Abundance of elements}

The spectra of the program stars are affected severely by line blending.
We have used a standard abundance analysis procedure based on the 
equivalent width measurements  only for those elements  for which we
could measure more than two clean lines.  
Spectral 
synthesis method is used for elements for which we could  measure only 
one good line. Local thermodynamic equilibrium is assumed for  the 
spectrum synthesis calculations. We have used the latest version of MOOG 
Sneden (1973) for spectrum synthesis.  The line list for each region  
synthesized is taken from the Kurucz atomic line list 
({\tt http://www.cfa.harvard.edu/\-amp/\-ampdata/\-kurucz23/\-sekur.html}) 
and from the Vienna Atomic Line 
Database \\ ({\tt http://ams.astro.univie.ac.at/vald/}).
Reference solar abundances for the various elemental species  are adopted from
Asplund, Grevesse \& Sauval (2005). The log\,gf values for atomic lines are
also  adopted from Fuhr et al. (1988) and Martin et al. (1988),
and  from a compilation of $gf$ values by R. E. Luck
(private communication). For heavy neutron-capture elements log\,gf values
given by Sneden et al. (1996) and Lawler et al. (2001) are   consulted.
The atomic parameters of the lines used for  abundance determination  of the 
elements  and their measured equivalent widths  are given in 
appendix  (Table 13).  The derived elemental abundances of the stars 
are presented in  Tables 7 through 10 .  In these Tables,  we have listed  
the abundance log $\epsilon$(X), along with [X/H] and  [X/Fe] values. 
In computing the quantity [X/Fe]  we have used the Fe~I-based abundance 
for elemental abundances derived from neutral lines and the Fe~II-based 
abundance for elemental abundances derived from ionized lines.
Elemental abundances  are discussed in the following sections.

\subsection{ Carbon, Nitrogen, Oxygen}
 
The  Carbon abundance is derived from spectral synthesis calculation 
of the G-band of CH around 4315 \AA\,.  A synthetic spectrum, derived with
an  appropriate model atmosphere  and using a carbon abundance of 
log ${\epsilon}$(C) = 7.7 ${\pm}$ 0.3, shows a good match to the depth 
of the observed spectrum of HD~26. Relative to the solar photospheric C
abundance, C is  mildly enhanced in HD~26 (${\rm [C/Fe]} = +0.31$ ). 
Spectrum synthesis calculation also shows  Carbon to be  strongly 
enhanced in HD~198269 and HD~224959 with [C/Fe] values of  2.30 and 2.01 
respectively. Spectrum synthesis fits of the region 4310 - 4330 \AA\, 
is shown in  Figure 1.
\begin{figure*}
\epsfxsize=14truecm
\epsffile{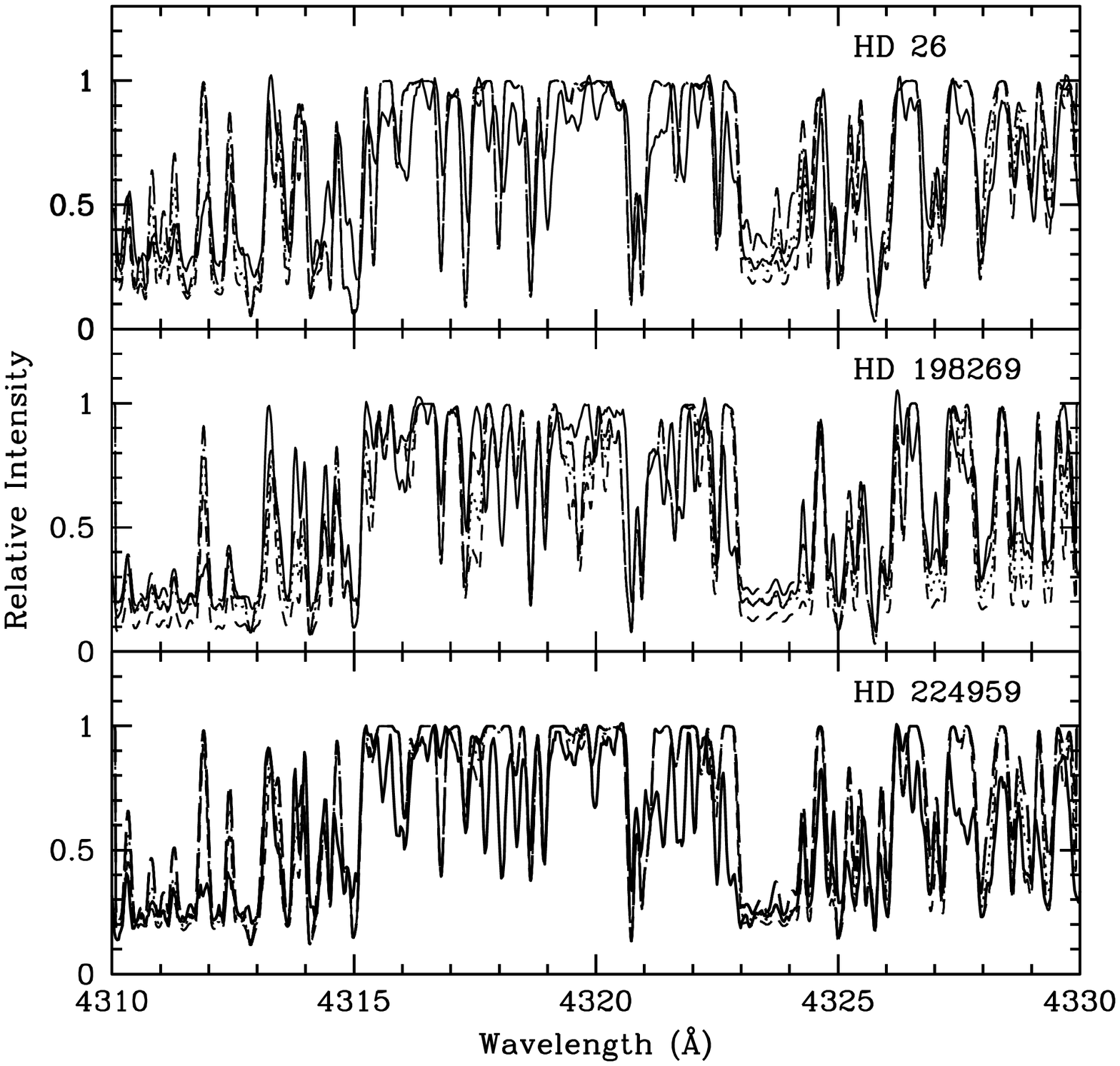}
\caption{ Spectrum synthesis fits of the region 4310 - 4330 \AA\,
 (dotted curve) compared with the observed spectra (solid curve) of
HD~26 (top panel).
Spectrum synthesis fits for HD~198269 and HD~224959
are shown in the middle  and the bottom panel respectively.
The synthetic
spectra are obtained using a model atmosphere corresponding to the adopted
parameters listed in Table 6. Two alternative synthetic spectra for
${\Delta}$[X/Fe] = +0.3 (long dash) and ${\Delta}$[X/Fe] = $-$0.3 (dot-dash)
are  shown to demonstrate the sensitivity of the line strength to
the abundances.}
%\label{Figure 1}
\end{figure*}

Reported C, N, O overabundances with respect to Fe ([X(C,N,O)/Fe]) in HD~26, 
and  HD~224959 (Masseron et al. 2010) are respectively
(0.68, 0.94, 0.36), (1.77, 1.88, 1.10). A carbon abundance of 
log ${\epsilon}$(C) of 7.45 is reported for HD~198269. 
The  Carbon  isotopic ratio ($^{12}$C/$^{13}$C = 5)  is  same as 
that of Vanture (1992b). Vanture (1992b) has determined oxygen 
abundance (log ${\epsilon}$(O)) for these objects  from
O I near-infrared triplet near  7774 \AA\, applying non-LTE corrections;
these  values are  ${\le}$ 7.7 for HD~224959, ${\le}$ 7.6 for HD~198269
and 8.45 for HD~26. With respect to Fe, Vanture  found carbon to be
enhanced by +0.45 (HD~26), +0.91 (HD~198269) and + 1.71 (HD~224959).
 
No clean oxygen lines are  detected in our spectra. We have made a rough
estimate of oxygen by considering the fact that, the CH stars for 
which oxygen abundances have been determined  follow a particular trend:  
Oxygen abundance in halo stars are found to increase with decreasing 
metallicity as [O/Fe] ${\sim}$ $-0.5$ [Fe/H] until [Fe/H] = -1.0 and 
then level off at a value of [O/Fe] = +0.35 ( Wheeler et al. 1989, 
Goswami \& Prantzos 2000). Based on this the Oxygen abundances 
(log$\epsilon$(O))  are respectively  7.9 (HD~26), 7.1 (HD~198269), 
6.6 (HD~224959) and 9.0 (HD~100764).

\subsection{The odd-Z elements Na and Al }

  Na abundance is  calculated from the resonance doublet - Na I D lines 
at $5890$\,{\AA} and $5896$\,{\AA}. These resonance lines are sensitive 
to non-LTE effects (Baum\"uller \& Gehren 1997; Baum\"uller et al. 1998; 
Cayrel et al. 2004). The non-LTE correction could be as high as 
$\pm$ 0.1 dex. The derived abundance  ${\rm [Na/Fe]}$ from an LTE analysis 
are  +0.29 (HD~26), +0.16 (HD~224959) and +0.20 (HD~198269).  HD~26 gives 
[Na/Fe] in marginal  excess of the First Dredge Up (FDU) predictions of  
El Eid \& Champagne (1995) ([Na/Fe] of +0.18 for a 5 M$_{\odot}$ 
stellar  model. Enrichment of Na is believed to result from  mixing of 
products  of Ne-Na cycle  involving  proton-capture on $^{22}$Ne in 
H burning region, following first dredge up.

 Al lines in our spectral coverage  are severely blended
 and could not be used for abundance determination.

\subsection{The ${\alpha}$-elements  Mg, Si, Ca, Ti }
{\it Magnesium} (Mg):
 The Mg abundance is derived from the synthesis of the
Mg~I line at 5172.68 {\AA}. The predicted line profile  with our 
adopted Mg abundances  fits the observed line profile  quite well. {\bf We have 
also measured the Mg abundance using lines at 4057.500, 4571.100 and 
5528.400 \AA. The presented abundance ratios in the Tables 7 - 10 are 
an average of the abundances derived from these four lines.} 
Magnesium is found to exhibit an overabundance  with  [Mg/Fe] $\sim +0.56$ 
in  HD~26.  For  HD~198269 and HD~224959 we have obtained [Mg/Fe] = +0.41 
and +0.25 respectively.

{\it Silicon} (Si):
 Lines of Si detected in the spectra of these objects are
severely blended and could not be used for abundance determination.
 
{\it Calcium} (Ca):
Abundance of Ca is derived using  three lines of  Ca I at 6102.72, 6122.22 
and 6439.07 \AA\,   for HD~26. Eight lines  for HD~198269 and
four for HD~224959 are used to derive Ca abundance.
The derived abundance estimates  are respectively
   ${\rm [Ca/Fe]}$ = +0.24 (HD~26), +0.52 (HD~198269) and +0.40 (HD~224959).
Estimated Non-LTE corrections for abundances derived from Ca I line 
 for dwarfs and subgiants is
about +0.1 dex. Hence we do not expect a large change in our estimates
of [Ca/Fe] caused by Non-LTE effect. Abundance of Ca 
could not be estimated for HD~100764 from our spectrum.

{\it Scandium} (Sc):  Spectrum synthesis of Sc II line
at 6245.63 \AA\, considering hyperfine structure from Prochaska and
McWilliam (2000) gave [Sc/Fe] = 0.52 for HD~198269, [Sc/Fe] = $-$0.20
for HD~26, and [Sc/Fe] = $-$0.89 for HD~224959. Spectrum synthesis
plots are shown in  Figure 2. Sc II line at 5031.02 \AA\, could be 
measured only for the star HD~198269. Due to the presence of C$_{2}$ lines
in this region, estimates of Sc abundace using this line would  likely 
to return inaccurate estimate. 
\begin{figure*}
\epsfxsize=12truecm
\epsffile{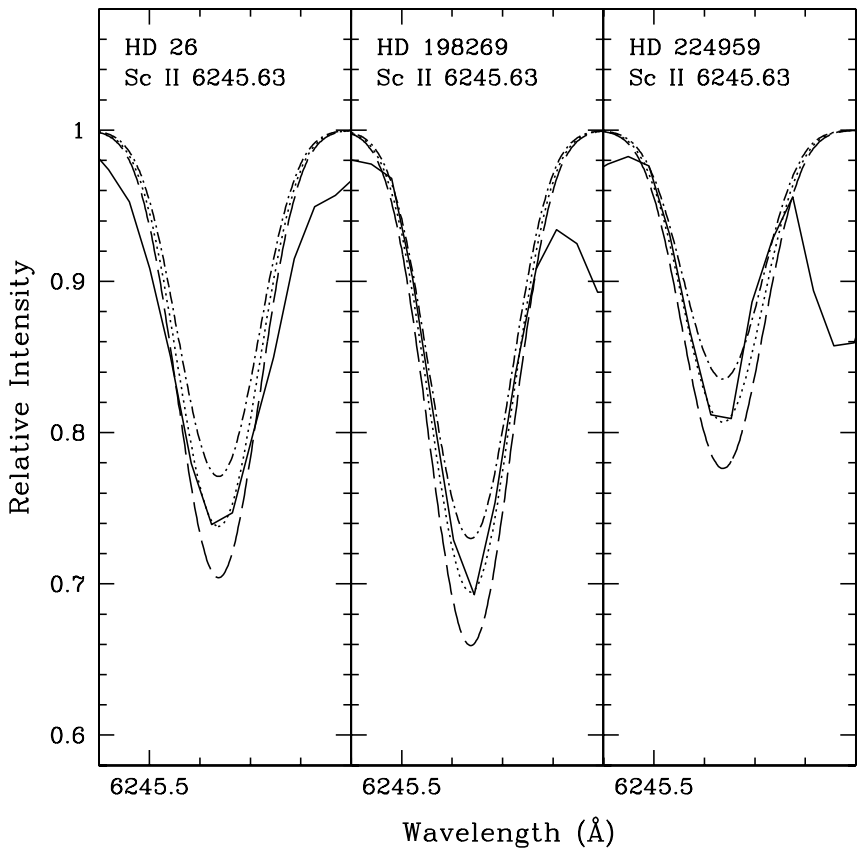}
\caption{Spectrum synthesis fits of Sc  (dotted curve)
compared with the observed spectra (solid curve) of HD~26, HD~198269 and 
HD~224959.  The synthetic
spectra are obtained using a model atmosphere corresponding to the adopted
parameters listed in Table 6. Two alternative synthetic spectra for
${\Delta}$[X/Fe] = +0.1 (long dash) and ${\Delta}$[X/Fe] = $-$0.1 (dot-dash)
are  shown to demonstrate the sensitivity of the line strength to
the abundances.}
%\label{Figure 2}
\end{figure*}

{\it Titanium} (Ti):  The  abundance of Ti  derived from Ti I and Ti II lines
differ slightly. For HD~26, two Ti I lines at 4533.24 and 4534.77 \AA\, and
four Ti II lines at 4418.33, 4443.794, 4563.77 and 4865.61 \AA\, are used. 
Log gf values are taken from Kurucz data base. 
Abundances with respect to Fe ([Ti I/Fe] and [Ti II/Fe])
are respectively  $-$0.06, +0.17  in HD~26, 0.22 and 0.48 in HD~198269, 
and 0.44, 0.49 in HD~224959. The number of Ti I and Ti II lines used for the
calculations are four and eight respectively for HD~224959 and seven and 
thirteen respectively for HD~198269. Non-LTE correction for the  abundance 
of Ti derived from Ti I  is about +0.1 dex  for dwarf and subgiants. If 
we consider this correction factor the difference between the abundances 
derived from Ti I and Ti II lines will reduce at most by 0.1 dex and the 
abundances obtained  from T I and Ti II lines  of HD~224959 would be 
almost similar. No lines of Ti were found usable  to derive Ti 
abundance for HD~100764.

\subsection{The iron-peak elements V, Cr, Mn, Ni, Zn}

{\it Vanadium} (V)\,:  Abundance of V in HD~198269 is estimated from
spectrum synthesis calculation of V I line at 5727.028 \AA\, taking
into account the  hyperfine components from Kurucz database that
gives [V/Fe] = +0.13. This line could not be detected in HD~26
and HD~224959.

{\it  Chromium} (Cr)\,:  
The Cr abundance derived using four Cr I lines  give [Cr/Fe] = $-$0.13 
for HD~198269.  The Cr abundance derived from a single Cr I line at 
4626.19 \AA\ returns a [Cr/Fe] value of  $-$0.19  for HD~26. Estimates of
[Cr/Fe]  for HD~198269 and   HD~224959 are respectively $-$0.08 
(five lines) and  $-$0.24 (two lines). 

{\it Manganese} (Mn)\,: The Mn abundance is derived from three Mn I lines
for HD~198269 give [Mn/Fe] = $-$0.13.  No clear good lines due to Mn  are  
detected in the spectra of HD~26, HD~224959 and HD~100764. 
The abundance of Mn derived using spectrum synthesis calculation of Mn I
line at 6013.51 \AA\, line taking account of hyperfine structures from 
Prochaska and McWilliam (2000) returns a value [Mn/Fe] = $-$0.13 for
HD~198269. This value for HD~26 is found to be $-$0.23.
The Mn I line at 6013.51 \AA\, could not be detected in HD~224959.

{\it Nickel} (Ni)\,:  Abundance of Ni in HD~26 is derived from two Ni I lines 
at 4980.16 \AA\, and 5035.36 \AA\,. Derived [Ni/Fe] = $-$0.06.
We could use only one good line of Ni I to derive its 
abundance in  HD~224959. [Ni/Fe] is found to be
 0.09.  For HD~198269 we have used five lines of Ni I
and derived a value $-$0.01 for [Ni/Fe].

{\it Zinc} (Zn)\,: The Zn abundance is derived from a single  Zn I line
at 4810.53 \AA\,  in HD~26.
[Zn/Fe] for HD~26 is found to be solar. No clear good lines could 
be detected in the spectra of HD~198269 and HD~224959. Estimated [Zn/Fe]
 is 0.11 for HD~100764.

\subsection{The light s-process elements Sr, Y, and Zr}

{\it Strontium} (Sr)\,: The Sr abundance is derived from Sr I line
at  $4607.327$~{\AA}.
Sr shows an overabundance  (${\rm [Sr/Fe]}$) of  1.89 for HD~26, 1.06 for
HD~198269 and 1.50 for HD~224959. 
Sr II lines at 4077.7 \AA\, and   4215.52  \AA\, are  blended, the 
latter  with  contributions from the CN molecular band around 4215 \AA\,.
Sr II line at  4161.82 \AA\,  could not be detected in the spectra of 
these objects.
Sr I line at 6550.24 \AA\, is   blended  with a Sc II line and therefore 
not used for abundance calculation. 
Sr abundance in HD~100764 could not be estimated; the spectral synthesis 
calculation of 
4607.327 \AA\, line did not give an acceptable synthetic fit with the
 observed feature in HD~100764.

{\it Yttrium} (Y)\,: The abundance of Y  with respect to Fe 
 shows an overabundance (${\rm [Y/Fe]}$) of  0.89 in HD~26, and 0.22 in both 
HD~198269 and  HD~224959. 
Among several lines of Y II examined in the spectra of these objects,
Y II 5200.41 \AA\, line appears as a good line. 
Y II line at 5205.73 \AA\, is also detected as a shallow feature.
The line at 4398.1 \AA\,
is asymmetric on the right wing, probably blended with a Nd II line at
4398.013 \AA\, and  could not be used for abundance determination.
Y II line at 4883.69 \AA\, is also asymmetric on the right wing,
blended with a Sm I line at 4883.777 \AA\,.  YII line  at 5087.43 \AA\, 
shows asymmetry on the right wing, and  5123.22 \AA\,  line appears  
as a shallow  feature, both the wings  not reaching the continuum. 
Y I line at 4982.129 \AA\, is blended with a Mn I line.  The  line
at 5473.388 \AA\, is blended with contributions from Mo I and Ce I lines.
Spectral synthesis calculation of Y II line at
$5200.41$\,{\AA} returns a near-solar value for HD~100764 with [Y/Fe] = $-$0.08.

{\it Zirconium} (Zr)\,: The  abundance of Zr is derived using  a single Zr II
 line
at $4317.321$ {\AA}.  Zr shows an overabundance  with ${\rm [Zr/Fe]}$ of 1.16
for HD~26.
Spectral synthesis of Zr I line
at 6134.57 \AA\, which is weakly detected in the spectrum of  HD~26 did 
not give a good synthetic fit with the observed 
feature and returned a value of [Zr/Fe] = 1.16.
None of the Zr I lines are usable for  abundance determination.
Zr II line at 4496.97 \AA\, appears as a blend with Co I
line at 4496.911 \AA\,. Absorption  tip of Zr II line at 4161.21 is easily
detected. Lines  detected at 4208.99, 4258.05 and 4317.32 \AA\, are
heavily contaminated by  contributions from molecular bands.
Zr II line at 4404.75 \AA\, is blended with  an Fe I line and  the
 4317.321 \AA\,  line is  blended with a  Ce II line. 
Spectrum synthesis fits of Sr, Y and Zr  are shown in  Figure 3
for HD~26.
Abundance of Zr  could not
be estimated for HD~100764 as no clean lines of Zr
could be detected in its spectrum. 
\begin{figure*}
\epsfxsize=12truecm
\epsffile{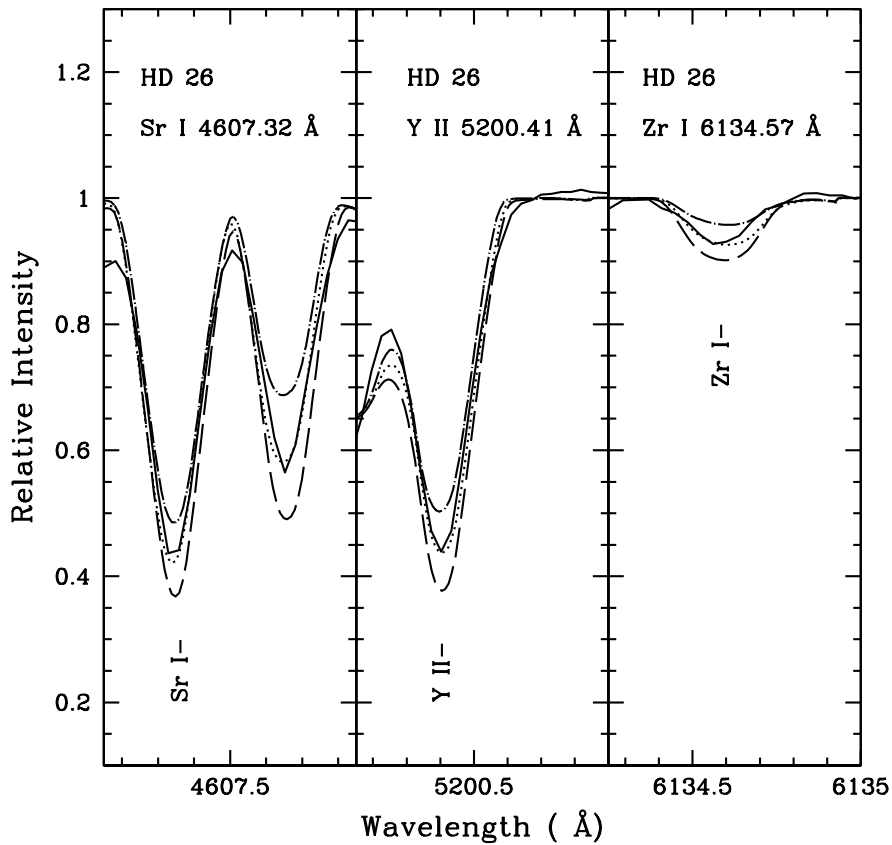}
\caption{Spectrum synthesis fits of Sr, Y and Zr  (dotted curve)
compared with the observed spectra (solid curve) of HD~26.  The synthetic
spectra are obtained using a model atmosphere corresponding to the adopted
parameters listed in Table 5. Two alternative synthetic spectra for
${\Delta}$[Sr, Y, Zr/Fe] = +0.3 (long dash) 
and ${\Delta}$ [Sr, Y, Zr/Fe] = $-$0.3 (dot-dash)
are  shown to demonstrate the sensitivity of the line strength to
the abundances.}
%\label{Figure 3}
\end{figure*}

\begin{figure*}
\epsfxsize=12truecm
\epsffile{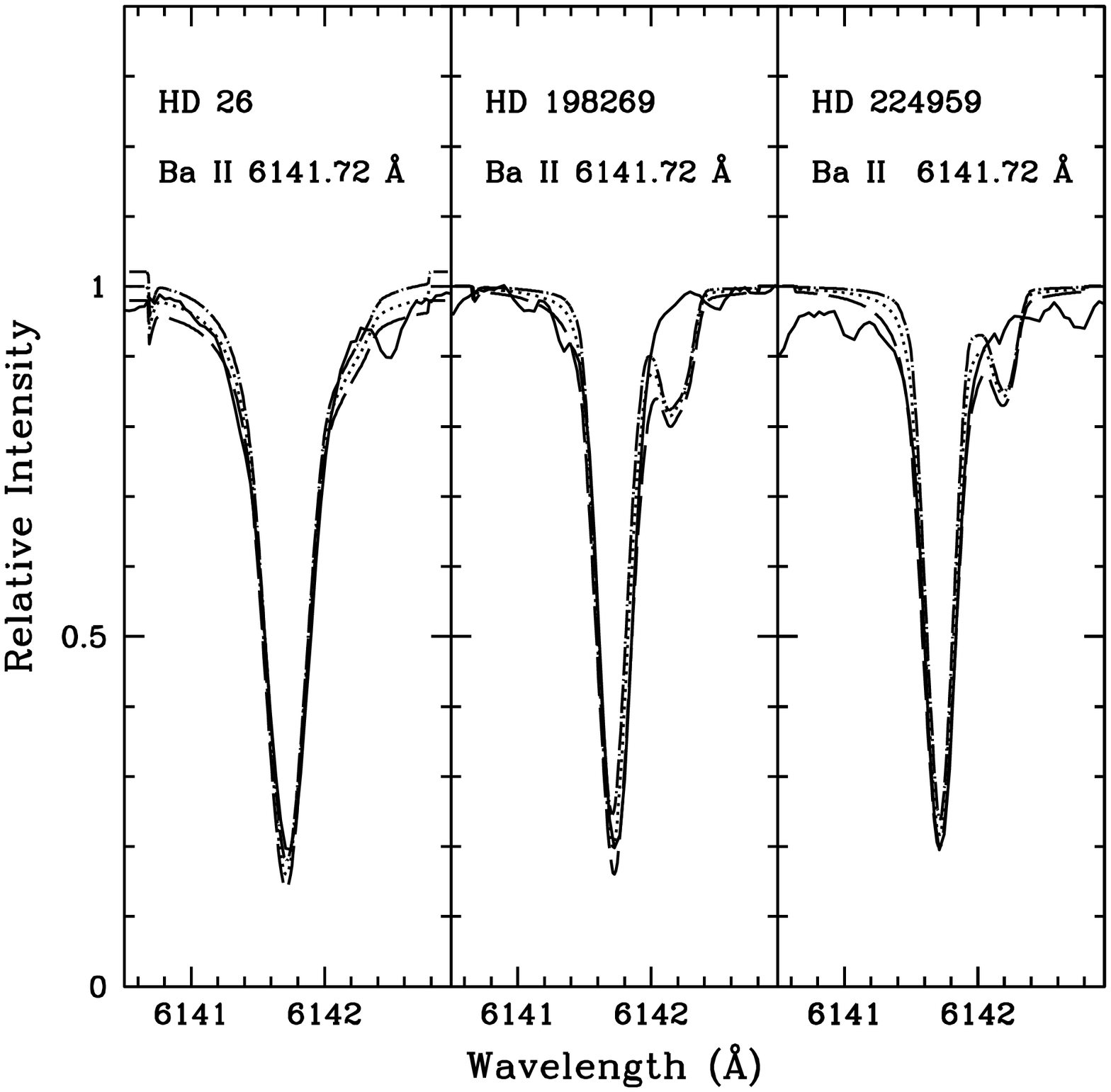}
\caption{ Spectrum synthesis fits of Ba  (dotted curve) compared
with the observed spectra (solid curve) of HD~26,
HD~198269 and HD~224959. The synthetic
spectra are obtained using a model atmosphere corresponding to the adopted
parameters listed in Table 5. Two alternative synthetic spectra for
${\Delta}$ [Ba/Fe] = +0.3 (long dash) and ${\Delta}$ [Ba/Fe] = $-$0.3 (dot-dash)
are  shown to demonstrate the sensitivity of the line strength to
the abundances. }
%\label{Figure 4}
\end{figure*}

\begin{figure*}
\epsfxsize=12truecm
\epsffile{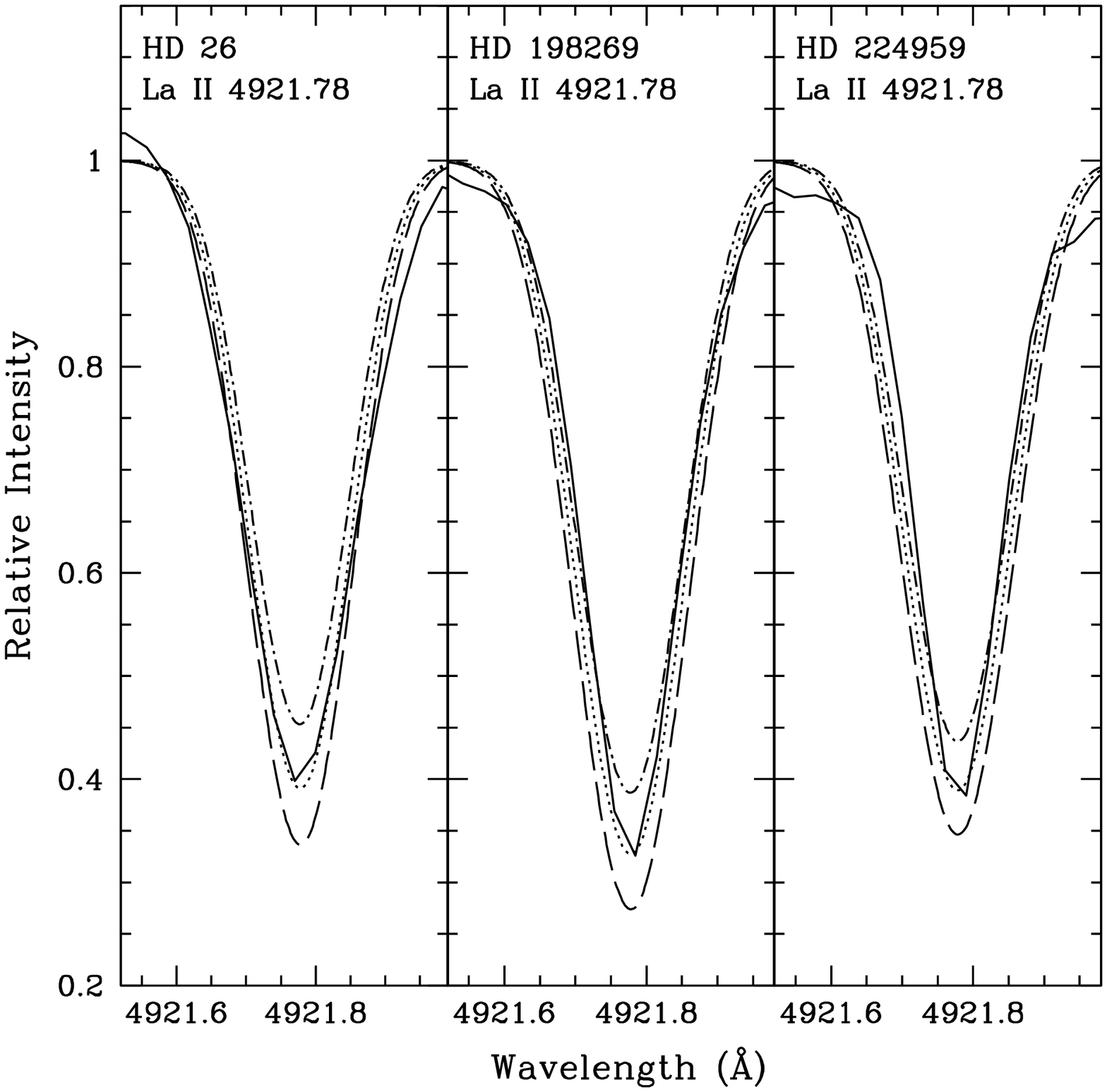}
\caption{ Spectrum synthesis fits of La  (dotted curve) compared
with the observed spectra (solid curve) of HD~26,
HD~198269 and HD~224959. The synthetic
spectra are obtained using a model atmosphere corresponding to the adopted
parameters listed in Table 6. Two alternative synthetic spectra for
${\Delta}$ [La/Fe] = +0.3 (long dash) and ${\Delta}$ [La/Fe] = $-$0.3 (dot-dash)
are  shown to demonstrate the sensitivity of the line strength to
the abundances. }
%\label{Figure 5}
\end{figure*}

\subsection{The heavy $n$-capture elements:  Ba, La, Ce, Pr, Nd,
Sm, Eu, Er, W, Pb}

The abundances of the heavy $n-$capture elements  barium (Ba),
lanthanum (La), cerium (Ce), neodymium (Nd),  samarium (Sm),
praseodymium (Pr), europium (Eu), erbium (Er), tungsten (W) and 
lead (Pb)  with respect to Fe  are generally found to be 
overabundant.  As the abundances  derived from weak lines are not 
expected to be affected by hyperfine splitting (McWilliam et al. 1995a, b), 
we have not used hyperfine-splitting corrections for  Y, Ce, and Nd lines.

{\it Barium} (Ba)\,:  The  Ba II line at 6141.727 \AA\, is used
 to determine the Ba abundance in HD~26.
Compared to  the other barium lines detected in the spectra this 
line appeared as a well defined  symmetric line in all the three 
CH stars spectra. log gf value for this line is taken from Miles 
and Wiese (1966). We have derived [Ba/Fe] =  1.93 for HD~26. Barium 
abundance determined using the lines at 5853.67, 6141.71 and 6496.90 
{\AA}\, returns a value  1.24 for [Ba/Fe]. The lines used for 
HD~224959 are  4934.08, 5853.67 and 6171.73 {\AA}\,; these lines return 
a value with [Ba/Fe] = 2.07 for this object.
 Abundance of Ba is found to 
be near-solar with [Ba/Fe] = -0.02 for HD~100764, using the spectrum 
synthesis calculation of 6141.727 \AA\,.
 Literature values of Ba  abundance for these objects are not available. 
The spectrum synthesis fits of Ba feature at 6141.7 \AA\, is 
shown in  Figure 4. Abundance of barium derived using spectrum synthesis
calculation using Ba II line at 5853.668 \AA\, considering hyperfine
components from McWilliam (1998) gave [Ba/Fe] = 2.75 for HD~224959,
 higher than what is obtained from 6141.7 \AA\, line. This line could not 
be used for  HD~26.

{\it Lanthanum} (La)\,:   
Three good lines of La are used for HD~26, two for HD~196289 and HD~100764 
each, and four for HD~224959  to derive La abundances. La exhibits an 
overabundance  (${\rm [La/Fe]}$) of  1.53 in HD~26, 1.57 in HD~198269, 
0.83 in HD~100764, and 2.50 in HD~224959. Our estimate of [La/Fe]  
matches closely with those of Van Eck et al. and Vanture for HD~224959 
and HD~198269.  For HD~26, our estimate ([La/Fe] = 1.56) is closer to 
Vanture's ([La/Fe] = 1.4), but about 0.8 dex lower than that given by 
Van Eck et al. ([La/Fe] = 2.3. No previous estimate  ([La/Fe]) for 
HD~100764  is available for a comparison.
Spectrum synthesis calculation of La II line at 4921.77 \AA\, (shown in Figure 5) considering 
hyperfine components from Jonsell et al. (2006) give [La II/Fe] = 1.0
for HD~26, 1.66 for HD~198269 and 2.6 for HD~224959.

{\it Cerium} (Ce)\,:
We have examined five Ce I lines and thirty three Ce II lines in our spectra.
A few of them are either affected by molecular contamination or blended with 
contributions from other elements. The abundances of Ce are derived using 
seven good lines of Ce II for HD~26,  twenty one for 
HD~198269 and seventeen for HD~224959. An overabundance  (${\rm [Ce/Fe]}$) 
of 1.65 is estimated for HD~26. This value agrees well with that of
Van Eck et al. and 0.28 dex lower than that of Vanture's. Our estimated [Ce/Fe]
for HD~198269  $\sim$ 1.66 is same as that obtained
by Vanture (1992c) and 0.1 dex higher than the estimate  of Van Eck et al.
 Our estimated [Ce/Fe] $\sim$ 2.28 for HD~224959 is about 0.3 dex higher
than that of Van Eck et al. and 0.15 dex higher than that of  Vanture's
estimate. We have derived [Ce/Fe] = 1.66 
 for HD~100764  using seven good Ce II lines. No previous estimate of [Ce/Fe] 
is available for this object.

{\it Praseodymium}  (Pr)\,: The abundance of Pr is derived using
four good lines each for HD~26 and HD~224959, and five lines are used 
for HD~198269. Pr exhibits an overabundance with ${\rm [Pr/Fe]} = +1.64$ 
in HD~26, 1.49 in HD~198269 and 2.35 in HD~224959. 
 Literature values of Pr  abundance for these objects are not available. 
Among the  one Pr I line and eight Pr II
lines examined in our spectra, Pr I line at $5996.060$\,{\AA} is detected 
as an asymmetric line.  Pr II line
at $5188.217$\,{\AA} is blended with a La II line and lines 
at $5219.045$\,{\AA} and $5220.108$\,{\AA} appeared as  asymmetric lines.
 As none of the Pr lines detected in the spectrum of HD~100764 is  usable 
for the abundance analysis, 
 Pr abundance could not be estimated  
for HD~100764.

{\it  Neodymium } (Nd)\,: One Nd I and thirty six Nd II lines listed in 
 Table 13  are examined. Most of them are blended
with contributions from other elements.
Twelve good lines of Nd II are used for HD~26, twenty one for HD~198269 and
twenty for HD~224959 to derive abundance of Nd in these objects.
 Neodymium  shows  an overabundance
 with ${\rm [Nd/Fe]}$ = 1.45 for HD~26. This value agrees with that
given by  Vanture (1992c) but higher by 0.15 dex than that estimated by  Van Eck
et al. (2003).
Luck \& Bond (1982) estimated [Nd/Fe] = 1.99 for this object.
[Nd/Fe] = 2.30 obtained for  HD~224959 and [Nd/Fe] = 1.48 obtained for
 HD~198269 are about  0.25 dex higher those of 
Van Eck et al.  (2003) and 0.45 dex higher than that of  Vanture (1992c).
Abundance of Nd  derived from a single line shows an overabundance with
[Nd/Fe] = 1.08 for HD~100764. 

{\it Samarium }  (Sm)\,: Two Sm I and eighteen  Sm II lines have been examined
in the spectra of these objects. The abundance of Sm is derived using five 
 Sm~II   lines for HD~26, seventeen for HD~198269 and eleven for HD~224959. 
Estimated  overabundance with ${\rm [Sm/Fe]}$ = 1.88
for  HD~26 is 0.6 dex higher than that of Van Eck et al. (2003) but matches well with
that of Vanture's estimate. Luck \& Bond (1982) gave a much lower value 
of [Sm/Fe] = 0.4 for this object.
In case of HD~198269 and 224959 our estimates
are  higher  than those of both Van Eck et al. and Vanture (Table 11).  
Abundance of Sm could not be derived for HD~100764.
One Sm I and two Sm II lines detected in the spectrum of HD~100764
are not usable for  abundance determination.

{\it Europium } (Eu)\,:
The abundance of Eu  is  determined for HD~26 and HD~224959 from a single 
good  line at $6437.64$\,{\AA} using spectral synthesis calculation.  Eu is
overabundant  (${\rm [Eu/Fe]}$) with 0.61 for HD~26 and 2.01 for HD~224959.
Previous estimates of Eu in these objects are not available. We could not 
estimate Eu abundance for HD~100764 and HD~198269 from our spectra as no 
clean lines could be detected. Spectrum synthesis calculation of 
$6437.64$\,{\AA} line did not give good synthetic fits. The blue Eu~II 
lines at 4129.7 and $4205.05$\, {\AA} and  the line at  $6645.13$\,{\AA} 
are severely blended  with strong molecular features  and could not be used 
for abundance analysis. 

{\it Erbium } (Er)\,:
The  abundance of Er  is estimated using
Er II line at  4759.671 {\AA}\,. The line parameters adopted from
Kurucz atomic line list come from Meggers et al. (1975). Er shows an
overabundance ([Er/Fe]) of 1.44 in HD~26, and 1.55 in HD~198269.
Er abundance could not be determined for HD~224959 and HD~100764
  from our spectra. Er II line at 4820.354 {\AA}\, is blended with
Ti II line at 4820.368 {\AA}\,.

{\it Tungsten } (W)\,: The abundance of W is determined using the W I line
at 4757.542 {\AA}\,. The line parameters  taken from the  Kurucz
atomic line list  come from Obbarius and Kock (1982). The abundance derived
is high  with [W/Fe] = 3.43 for HD~26 and 2.75 for HD~198269; there is a 
possible  blend with Cr I line at
4757.578 \AA\,  (with log g and lower excitation potential, -0.920 and 3.55
respectively) resulting  in an  overestimate of  W abundance. For 
HD~224959 and 100764  abundance of W could not be estimated from our spectra.

{\it Osmium } (Os)\,: Three lines of Os I are detected. Os I at 4048.054 {\AA}\,
is detected in  the spectra of all the three stars, HD~26, HD~198269 
and HD~224959. However, this line has a possible blend with Cr II line 
at 4048.025 {\AA}\, and could not be used for abundance estimates. The other
two lines at 4793.996 {\AA}\, and 4865.607 {\AA}\, could be detected 
only in the spectrum of HD~198269. These two lines are blended with
Ca I line at 4794.015 {\AA}\, and Ti II line at 4865.612 {\AA}\,
respectively. The line at 4793.996 {\AA}\, returns a value showing
enhancement with [Os/Fe] = 1.67 in HD~198269.

 {\it Lead } (Pb)\,:  Spectrum-synthesis calculation is also used
to determine the abundance of Pb  using the Pb I line at
$4057.8$\,{\AA}.  Pb shows an overabundance
 (${\rm [Pb/Fe]}$) of  2.11 for HD~26, 2.4 for HD~198269 and 
3.70 for HD~224959. 
This line is strongly affected by molecular absorption of
CH. CH lines are included in our spectrum
synthesis calculation.
This line could not be detected in the
spectrum of HD~100764. Our estimated [Pb/Fe] for HD~198269 is similar 
to the estimate of Van Eck et al. 2003. For HD~26 and HD~224959 our 
estimated [Pb/Fe] is
about 0.5 dex higher than that obtained by Van Eck et al..
 The spectrum synthesis fits of the lead
feature is shown in  Figure 6 for HD~224959.
\begin{figure*}
\epsfxsize=13truecm
\epsffile{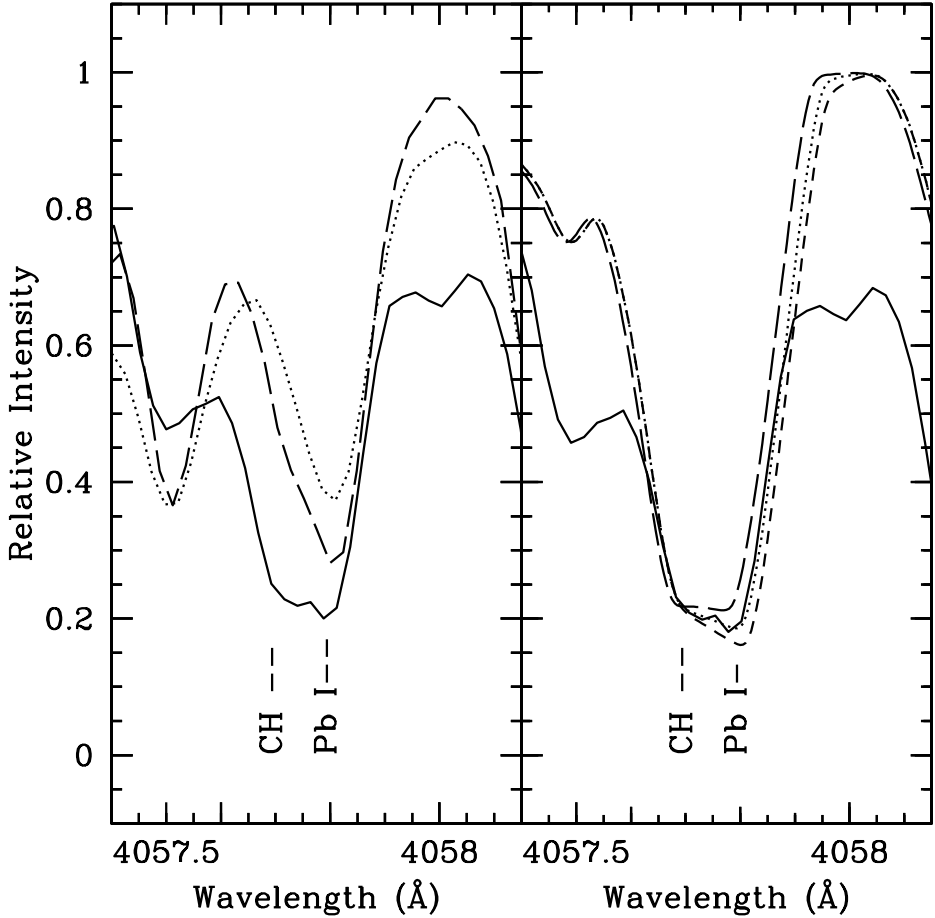}
\caption{ In the left panel the region around the  lead feature at
4057.8 \AA\, is compared in  the spectrum of HD~26 (dotted line),
HD~198269 (dashed line) and HD~224959 (solid line). In the right panel,
the  spectrum synthesis fits of Pb  (dotted curve) is  shown
with the observed spectra (solid curve) of HD~224959.
 The synthetic
spectra are obtained using a model atmosphere corresponding to the adopted
parameters listed in Table 6. Two alternative synthetic spectra for
${\Delta}$ [Pb/Fe] = +0.3 (long dash) and ${\Delta}$ [Pb/Fe] = $-$0.3
 (dot-dash)
are  shown to demonstrate the sensitivity of the line strength to
the abundances. }
%\label{Figure 6}
\end{figure*}

\subsection{Error Analysis}
Error analysis is discussed in detail in one of our earlier paper 
Goswami et al. (2006). Here we mention it briefly.
Random errors and systemic errors are the two main sources of errors
affecting the derived abundances.
 Random errors arise  due to uncertainties
in line parameters, i.e.  the adopted $gf$ values and the equivalent width
measurements. These  errors  cause line-to-line scatter in derived
abundances for a given species.   Random errors are  minimized by
employing as many usable lines as possible for a given element. In deriving
the Fe abundances we made use of Twenty two Fe I lines and four
Fe II lines in HD~26, twenty nine Fe I lines and eight
Fe II lines in HD~198269 and  twenty five Fe I lines and 9
Fe II lines in HD~224959. For the object  HD~100764  we could use nineteen Fe I and three Fe II lines. The
derived standard deviation ${\sigma}$ is defined by ${\sigma}^{2}$ =
[${\Sigma}(\chi_{i} - {\chi})^{2}/(N-1)$], where $N$ is the number of lines
used. The values of ${\sigma}$ computed from the Fe I lines are respectively 
${\pm}$ 0.24, 0.15, 0.24 and 0.27.
dex in HD~26, 198269, 224959 and 100764. The corresponding value
calculated for Fe II lines are respectively ${\pm}$ 0.18, 0.13, 0.27 
and 0.13.
 The computed errors for Fe I and Fe II are listed in the Tables 7 - 10.

We have calculated the error in the atmospheric parameters as described 
by Ryan et al. (1996).  The error in the atmospheric parameters have 
contibutions from the measurement of equivalent widths and inaccurate 
log gf values of Fe lines used for the analysis.  Since we have used 
the most updated log gf values  as given in Table 13, the contribution 
to the error from these values are negligible. To find the minimum error 
in the atmospheric parameters, we have used the standard deviation of 
the Fe abundance derived for each objects as listed in Table 7-10, which 
is found to be approximately 0.2 dex in all the objects. Along with this 
corresponding to 0.2 dex in abundance we have calculated the minimum 
error in the temperature as 100 K, log g as 0.03 and 0.06 km s$^{-1}$ 
in micro turbulent velocity. Then the minimum contribution from the 
stellar atmospheric parameters to the elemental abundances are calculated 
using the general equation

\begin {center}

$E_r$ = $\sqrt{{E_{r1}}^2 + {E_{r2}}^2 + {E_{r3}}^2 + {E_{r4}}^2+ {E_{r5}}^2......+ {E_{rn}}^2}$

\end{center}

The values are found to be 0.25, 0.16, 0.25 and 0.29 in HD 26, HD 198269, 
HD 224959 and HD 100764 respectively.

For abundances derived by spectrum synthesis calculations  we have 
visually estimated the fitting errors. The estimated fitting errors 
range between 0.1 dex and 0.3 dex. We adopt these fitting errors as 
estimates of the random errors associated with the derived elemental 
abundances. 

\section{Parametric model based analysis of the observed abundances}

We have examined the relative  contributions from  the s- and
r-process to  the observed abundances  by comparing the observed 
abundances with predicted s- and r-process contributions in the 
framework of a parametric model for s-process (Howard et al. 1986). 
We have used the solar system isotopic abundances (both s- and r-process) 
given in   Arlandini et al. (1999). The solar system  r- and s- process 
elemental abundances derived  from the isotopic abundances are scaled 
to  the  metallicity of the star.  These values are  normalized to the observed Ba abundance of the corresponding stars.
Using a  parametric model function
 $ N_{i}(Z) = A_{s}N_{s,i} + A_{r}N_{r,i}$,

where Z is the metallicity and  N$_{s,i}$ and N$_{r,i}$ are  the 
ith element abundance produced  respectively by s-process and  r-process.
The coefficients A$_{s}$ and  A$_{r}$ that represent the contributions 
coming from s- and r-process respectively are obtained from  non-linear 
least square fits. The contributions of r- and s-process elements are 
estimated from the fitting for Ba-Eu. The 1st peak elements Sr, Y and 
Zr are not included in the fitting as they are not  abundantly produced
as 2nd peak ones by the s-process in metal-poor AGB stars in general.
The model fits  are shown in  Figure 7.
The  derived coefficients A$_{s}$,  A$_{r}$ and  the reduced chisquare
($\chi^{2}_{\nu}$) values for HD~26, 198269 and 224959  are listed in Table 12.

{\footnotesize
\begin{table*}
{\bf Table 6: Derived  atmospheric parameters }\\
%\tiny
\begin{tabular}{|c|c|c|c|c|c|c|}
\hline
      &          &         &         &        &          &        \\
~~Star Names~~~  &~~~$T_{\rm eff}$~~~ &~~~log\,g~~~ &~~~ $V_{\rm t}$ km s$^{-1}$~~~ &~~~[Fe I/H]~~~& ~~~[Fe II/H]~~~ & ~~~Remarks~~~\\
               &  (K)       &      &     &          &        &      \\
\hline
               &         &      &     &          &        &      \\
HD~26          & 5000   & 1.6  & 2.07 & $-$1.11  & $-$1.15 & 1 \\ 
               & 5170   &  2.2 &     &          &         & 2 \\
               &         &      &     &  $-$0.4         &$-$0.4         & 4 \\
               &         &      &     &  $-$0.3         &$-$0.3         & 4  \\
               &         &      &     &  $-$1.25$\pm$0.3 &$-$1.25$\pm$0.3 & 5\\
               &         &      &     &  $-$0.45$\pm$0.4 &$-$0.45$\pm$0.4 & 6  \\
               & 5250    &  2.5 &     &  $-$0.45 &        &   10  \\
               &         &      &     &          &        &      \\
\hline
               &         &      &     &          &        &      \\
HD~100764      & 4750   & 2.0  & 2.0 & $-$0.82  & $-$0.90 & 1 \\
               & 4600   &      &     &          &         & 7  \\ 
               & 4850   &      &     &          &         & 8 \\
               & 4850   & 2.2  &  5.0 &   $-$0.6 &        &  9     \\
               &         &      &     &          &        &      \\
\hline
               &         &      &     &          &        &      \\
HD~198269      & 4500    & 1.5  & 2.25 & $-$2.03  & $-$2.05  & 1 \\
               & 4520   &      &     &          &         & 8 \\
               & 4800   &  1.3  &     &          &         & 2 \\
               &         &      &     &  $-$2.2$\pm$0.2 &$-$2.2$\pm$0.2 & 5  \\
               &         &      &     &  $-$1.4         &$-$1.4         & 4  \\
               & 4460    &      &     &  $-$1.56        &        &  11  \\
               &         &      &     &          &        &      \\
\hline
               &         &      &     &          &        &      \\
HD~224959      & 5050    & 2.1  & 2.0 & $-$2.42  & $-$2.42 & 1 \\
               & 5200    &  1.9 &     &   $-$1.6       &         & 2, 4  \\
               &         &      &     &  $-$2.2$\pm$0.2 &$-$2.2$\pm$0.2 & 5 \\
               &         &      &     &  $-$1.6         &$-$1.4         & 4 \\
               &         &      &     &          &        &      \\
\hline
\end{tabular}

References: 1: this work,  2: Vanture (1992a), 3: Vanture (1992b), 4: Vanture (1992c),\\
 5: Van Eck et al. (2003), 6: Luck \& Bond (1982), 7: Bergeat et al. (2001) \\
 8: Aoki \& Tsuji (1997), 9: Dominy, J. F. (1984)\\
10: Smith, V. V. \& Lambert, D. L. (1986); 11: Lee P., (1974)\\

\end{table*}
}

\clearpage
{\footnotesize
\begin{table*}
{\bf Table 7 : Elemental abundances in HD~26}\\
\begin{tabular}{|c|c|c|c|c|c|}
\hline
 &       &                         &                &        &        \\
 &    Z  &   solar $log{\epsilon}^a$ & $log{\epsilon}$& [X/H]  & [X/Fe] \\
 &       &                         &                  &        &        \\
\hline
C           &   6     &   8.39   &   7.7              & $-$0.69    &  +0.31  \\
N           &   7     &   7.78   &  7.9$^{b}$         &  $-$0.02   &  +1.12  \\
O           &   8     &   8.66   &  7.9               &  $-$0.76   & +0.34  \\
Na {\sc i}  &  11     &   6.17   &  5.35              &   $-$0.82  &  +0.29 \\
Mg {\sc i}  &  12     &   7.53   &  6.98              &   $-$0.55  &  +0.56  \\
Ca {\sc i}  &  20     &   6.31   &  6.44$\pm$0.11(3)  &   $-$0.87  &  +0.24\\
Ti {\sc i}  &  22     &   4.90   &  3.73$\pm$0.06(2)  &   $-$1.17  &  -0.06\\
Ti {\sc ii} &  22     &   4.90   &  3.96$\pm$0.30(4)  &   $-$0.94  &  +0.17\\
Cr {\sc i}  &  24     &   5.64   &  4.34(1)           &   $-$1.30  &  $-$0.19\\
Mn {\sc i}  &  25     &   5.39   &  4.05(1)           &   $-$1.34  &  $-$0.23\\
Fe {\sc i}  &  26     &   7.45   &  6.34$\pm$0.21(20) &   $-$1.11  &  -\\
Fe {\sc ii} &  26     &   7.45   &  6.30$\pm$0.16(4)  &   $-$1.15  &  -\\
Ni {\sc i}  &  28     &   6.23   &  5.06(1)           &   $-$1.17  &  $-$0.06\\
Zn {\sc i}  &  30     &   4.60   &  3.49(1)           &   $-$1.11  &  0.0  \\
Sr {\sc i}  &  38     &   2.92   &  3.7(1)            &   +0.78    & +1.89\\
Y {\sc ii}  &  39     &   2.21   &  1.95(1)           &   $-$0.26  &  +0.89\\
Zr {\sc i}  &  40     &   2.59   &  -                 &   -        & -\\
Zr {\sc ii} &  40     &   2.59   &  2.6(1)            &   +0.01    & +1.16\\
Ba {\sc ii} &  56     &   2.17   &  2.95(1)           &   +0.78    & +1.93\\
La {\sc ii} &  57     &   1.13   &  1.51$\pm$0.05(3)  &   +0.38    & +1.53\\
Ce {\sc ii} &  58     &   1.58   &  2.08$\pm$0.13(5)  &   +0.50    & +1.65\\
Pr {\sc ii} &  59     &   0.71   &  1.20$\pm$0.17(4)  &   +0.49    & +1.64\\
Nd {\sc ii} &  60     &   1.45   &  1.75$\pm$0.10(8)  &   +0.30    & +1.45\\
Sm {\sc ii} &  62     &   1.01   &  1.74$\pm$0.30(5)  &   +0.73    & +1.88\\
Eu {\sc ii} &  63     &   0.52   &  0.1(1)            &   $-$0.42  & +0.61\\
Er {\sc i}  &  68     &   0.93   &    -               &     -      &   - \\
Er {\sc ii} &  68     &   0.93   &  1.22(1)           &   +0.29    & +1.44 \\
W {\sc i}:   &  74     &   1.11  &  3.50(1)           &   +2.39    & +3.54 \\
Pb          &  82     &   2.00   &  3.0(1)            &   +1.0     & +2.11\\
\hline 
\end{tabular}

$^{a}$ Asplund et al. (2005), $^{b}$ Vanture (1992b) \\
Numbers within parenthesis in column 4 indicate the number of lines used for abundance estimate.\\
\end{table*}
}

{\footnotesize
\begin{table*}
{\bf Table 8 : Elemental abundances in HD~100764}\\
\begin{tabular}{|c|c|c|c|c|c|}
\hline
 &       &                                   &                &        &        \\
 &    Z  &   solar $log{\epsilon}^a$ & $log{\epsilon}$& [X/H]  & [X/Fe] \\
 &       &                           &                &        &        \\
\hline
Fe {\sc i} &  26     &   7.45   &  6.63$\pm$0.27(19) &   $-$0.82  &  -\\
Fe {\sc ii}&  26     &   7.45   &  6.55$\pm$0.13(3)  &   $-$0.90  &  -\\
Zn         &  30     &   4.60   &  2.21(1)           &   $-$0.71  & +0.11\\
Y {\sc ii} &  39     &   2.21   & 1.23(1)            &   $-$0.98  & $-$0.08\\
Ba {\sc ii} & 56     &   2.17   & 1.25(1)            &   $-$0.92  & $-$0.02 \\
La {\sc ii} &  57    &   1.13   &  1.06$\pm$0.02(2)  &   $-$0.07  & +0.83\\
Ce {\sc ii} &  58    &   1.58   &  2.34$\pm$0.02(7)  &   +0.76  & +1.66\\
Nd {\sc ii} &  60    &   1.45   &  1.63(1)           &   +0.18  & +1.08\\
\hline 
\end{tabular}

$^{a}$ Asplund et al. (2005)\\
Numbers within parenthesis in column 4 indicate the number of lines used 
for abundance estimate.\\
\end{table*}
}

\clearpage
{\footnotesize
\begin{table*}
{\bf Table  9 : Elemental abundances in HD~198269}\\
\label{tab:1}       
\begin{tabular}{|c|c|c|c|c|c|}
\hline
 &         &                     &                &        &        \\
 &    Z  & solar $log{\epsilon}^a$ & $log{\epsilon}$& [X/H]  & [X/Fe] \\
 &         &                         &                &        &        \\
\hline
C          &   6    &   8.39   &  8.9               &  +0.41     &  +2.30   \\
N          &   7    &   7.78   &  7.6$^{b}$         &   $-$0.18  &  +1.71   \\
O          &   8    &   8.66   & 7.1                &  $-$1.56   &  +0.33   \\
Na {\sc i}  &  11   &   6.17   &  4.30              &  $-$1.87   &  +0.16   \\
Mg {\sc i}  &  12   &   7.53   &  6.03$\pm$0.09(3)  &   $-$1.50  &  +0.53\\
Ca {\sc i}  &  20   &   6.31   &  4.78$\pm$0.14(8)  &   $-$1.53  &  +0.52\\
Sc {\sc ii} &  21   &   3.05   &  1.51(1)           &   $-$1.54  &  $+$0.51\\
Ti {\sc i}  &  22   &   4.90   &  3.09$\pm$0.14(7)  &   $-$1.81  &  +0.22\\
Ti {\sc ii} &  22   &   4.90   &  3.33$\pm$0.22(13) &   $-$1.57  &  +0.48\\
V  {\sc ii} &  23   &   4.00   &  2.1(1)            &   $-$1.9   &  +0.13 \\
Cr {\sc i}  &  24   &   5.64   &  3.53$\pm$0.18(5)  &   $-$2.11  &  $-$0.08\\
Mn {\sc i}  &  25   &   5.39   &  3.23$\pm$0.16(3)  &   $-$2.16  &  $-$0.13\\
Fe {\sc i}  &  26    &   7.45   &  5.42$\pm$0.17(29)&   $-$2.03  &  -\\
Fe {\sc ii} &  26    &   7.45   &  5.40$\pm$0.13(8) &   $-$2.05  &  -\\
Ni {\sc i}  &  28   &   6.23   &  4.19$\pm$0.14(5)  &   $-$2.04  &  $-$0.01\\
Sr {\sc i}  &  38    &   2.92   &  1.95(1)          &   $-$0.97  &  +1.06\\
Y {\sc ii}  &  39    &   2.21   &  0.38(1)          &   $-$1.83  &  +0.22\\
Ba {\sc ii} &  56   &   2.17   &  1.36$\pm$0.06(3)  &   $-$0.81  & +1.24\\
La {\sc ii} &  57   &   1.13   &  0.65$\pm$0.14(2)  &   $-$0.48  & +1.57\\
Ce {\sc ii} &  58   &   1.58   &  1.19$\pm$0.15(10) &   $-$0.39  & +1.66\\
Pr {\sc ii} &  59   &   0.71   &  0.15$\pm$0.09(4)  &   $-$0.56  & +1.49\\
Nd {\sc ii} &  60   &   1.45   &  0.88$\pm$0.11(11) &   $-$0.57  & +1.48\\
Sm {\sc ii} &  62    &   1.01   &  0.84$\pm$0.13(12) &   $-$0.17 & +1.88\\
Eu {\sc ii} &  63    &   0.52   &   -                &     -     &   -\\
Er {\sc ii} &  68    &   0.93   &  0.43(1)           &   $-$0.50 & +1.55\\
W {\sc i}   &  74    &   1.11   &  1.94(1)           &   +0.83   & +2.88\\
Os {\sc i}  &  76   &   1.45   &  1.07(1)            &   $-$0.38 & +1.67\\
Pb {\sc i}  &  82   &   2.00   &   2.5(1)            &   +0.5    &   +2.4   \\
\hline 
\end{tabular}

$^{a}$ Asplund et al. (2005),  $^{b}$ Vanture (1992b)\\
Numbers within parenthesis in column 4 indicate the number of lines used for abundance estimate.\\
\end{table*}
}

\clearpage
{\footnotesize
\begin{table*}
{\bf Table 10 : Elemental abundances in HD224959}\\
\begin{tabular}{|c|c|c|c|c|c|}
\hline
  &      &                &                &        &        \\
 &    Z  &   solar $log{\epsilon}^a$ & $log{\epsilon}$& [X/H]   & [X/Fe] \\
  &      &                           &                &         &        \\
\hline
C          &   6    &   8.39   &  8.0               &  $-$0.39   & +2.01   \\
N          &   7    &   7.78   &  8.2$^{b}$         &  +0.42     & +1.98   \\
O          &   8    &   8.66   &  6.6               &  +2.06     &  +0.34  \\
Na {\sc i}  &  11   &   6.17   &  3.95              &   $-$2.22  &  +0.20   \\
Mg {\sc i}  &  12   &   7.53   &  5.50              &   $-$2.03  &  +0.39  \\
Ca{ \sc i}  &  20   &   6.31   &  4.29$\pm$0.06(4)  &   $-$2.02  &  +0.40 \\
Ti {\sc i}  &  22   &   4.90   &  2.92$\pm$0.15(4)  &   $-$1.98  &  +0.44 \\
Ti {\sc ii} &  22   &   4.90   &  2.97$\pm$0.21(8)  &   $-$1.93  &  +0.49 \\
Cr {\sc i}  &  24   &   5.64   &  2.98$\pm$0.16(2)  &   $-$2.66  &  $-$0.24\\
Fe {\sc i}  &  26   &   7.45   &  5.03$\pm$0.15(21) &   $-$2.42  &  -\\
Fe {\sc ii} &  26   &   7.45   &  5.03$\pm$0.23(8)  &   $-$2.42  &  -\\
Ni {\sc i}  &  28   &   6.23   &  3.90(1)           &   $-$2.33  &  +0.09\\
Sr {\sc i}  &  38   &   2.92   &  2.0(1)            &   $-$0.92  &  +1.50\\
Sr {\sc ii} &  38   &   2.92   &  -                 &   -        &  -\\
Y {\sc i}   &  39   &   2.21   &  -                 &   -        &  -\\
Y {\sc ii}  &  39   &   2.21   &  0.01(1)           &   $-$2.2   &  +0.22\\
Ba {\sc ii} &  56   &   2.17   &  1.82$\pm$0.12(3)  &   $-$0.35  &  +2.07\\
La {\sc ii} &  57   &   1.13   &  1.21$\pm$0.05(2)  &   +0.08    &  +2.50\\
Ce {\sc ii} &  58   &   1.58   &  1.44$\pm$0.17(8)  &   $-$0.14  &  +2.28\\
Pr {\sc ii} &  59   &   0.71   &  0.64$\pm$0.11(3)  &   $-$0.07  &  +2.35\\
Nd {\sc ii} &  60   &   1.45   &  1.33$\pm$0.09(10) &   $-$0.12  &  +2.30\\
Sm {\sc ii} &  62   &   1.01   &  1.36$\pm$0.25(8)  &   +0.35    &  +2.07\\
Eu {\sc ii} &  63   &   0.52   &  0.09(1)           &   $-$0.43  &  +2.01\\
Pb {\sc i}  &  82   &   2.00   &  3.3(1)            &   +1.3     &  +3.70 \\
\hline 
\end{tabular}
\\
$^{a}$ Asplund et al. (2005),  $^{b}$ Vanture (1992b)\\
Numbers within parenthesis in column 4 indicate the number of lines used for abundance estimate.\\
\end{table*}
}

\clearpage
{\footnotesize
\begin{table*}
{\bf Table 11: Abundance ratios} \\
\tiny
\begin{tabular}{|c|c|c|c|c|c|c|c|c|c|}
\hline
         &         &        &       &      &          &             &      &                &       \\
Star &   [Fe I/H] & [Fe II/H] &[Fe/H] & [Sr/Fe]& [Y/Fe]  & [Zr/Fe]& [Ba/Fe]& [La/Fe]&  Ref\\
         &         &        &       &      &          &             &      &                &       \\
\hline
         &         &        &       &      &          &             &      &                &       \\
HD~26     &  $-$1.11  &  $-$1.15 &  $-$1.13&  1.89&  0.89 & 1.16       &  1.93 &  1.53             &   1\\
         &   -     &  -     &  $-$1.25&   -  &    -  & 0.9$\pm$0.3&   -   &   $2.3_{-0.5}^{+0.1}$ &   2\\
         &    -    &  -     &  $-$0.4 &    - &  1.0$\pm$0.03 &0.9$\pm$0.3 & - & 1.4$\pm$0.4   &   3\\
         &   -     &   -    &  $-$0.3 &   -  &    -     &   -         & -  &   -              &   4\\
         &   -     &   -    & $-$0.45 &    - &   -      & 0.6$\pm$0.41&  -  & 2.27$\pm$0.34   &   5\\
         &         &        &       &      &          &             &      &                &       \\
\hline
HD~100764 &  $-$0.82  &  $-$0.90 & $-$0.86 &   -  &  $-$0.08   &  -        & $-$0.02 & +0.83$\pm$0.02 &  1\\
         &         &        &       &      &          &             &      &                &       \\
\hline
HD~198269 &  $-$2.03  &  $-$2.05 &  $-$2.04&  1.06&  0.22    & --    &  1.24    &  1.57$\pm$0.14 & 1\\
         &    -    &   -    &  $-$2.2$\pm$0.2&   -  &    -    & 0.4$\pm$0.1&   -      &   1.6$\pm$0.2 & 2\\
         &    -    &    -   &  $-$1.4 &    - &   -   &1.2$\pm$0.1 & -                 & 1.4$\pm$0.4   &  3\\
         &         &        &       &      &          &             &      &                &       \\
\hline
HD~224959 &  $-$2.42 &  $-$2.42&  $-$2.42 &  1.50&  0.22     & --    &  2.07$\pm$0.12 &  2.50$\pm$0.08 &  1\\
         &    -    &    -   &  $-$2.2$\pm$0.2&   -  &    -    & 1.0$\pm$0.1&   -           &   2.3$\pm$0.2 &  2\\
         &    -    &    -   &  $-$1.6 &    - &   -   &- & -                       & 2.0$\pm$0.6   &   3\\
         &         &        &       &      &          &             &      &                &       \\
\hline 
\end{tabular}

1.This work, 2. Van Eck et al. (2003), 3. Vanture (1992c), 4. Thevenin and 
Idiart (1999), 5. Luck and Bond (1982)\\
\end{table*}
}

{\footnotesize
\begin{table*}
{\bf Table 11: Abundance ratios (Continued)} \\
\tiny
\begin{tabular}{|c|c|c|c|c|c|c|c|c|c|c|}
\hline
         &        &       &      &         &          &   &    &    &   &     \\
Star & [Ce/Fe]& [Pr/Fe]  & [Nd/Fe]& [Sm/Fe]& [Eu/Fe]& [Er/Fe]&[W/Fe]&[Os/Fe]&[Pb/Fe]&Ref\\
         &        &       &      &         &          &   &    &    &   &     \\
\hline
         &        &       &      &         &          &   &    &    &   &     \\
HD~26     &  1.65$\pm$0.13 & 1.64$\pm$0.17 & 1.45$\pm$0.10 & 1.88$\pm$0.30& 0.68 &   1.44 & 3.54 &    --  & 2.55$\pm$0.30  & 1\\
         & $1.7_{-0.4}^{+0.3}$ &    -   & $1.3_{-0.1}^{+0.3}$ &   1.3$\pm$0.3  &    -  & - &   -   &   - &  2.0$\pm$0.2 &  2\\
         & 1.9$\pm$0.4  &   -    & 1.6$\pm$0.15 & 1.9$\pm$0.2 &  - &-         & - & -  &  -&  3\\
         &        &       & 1.99 &    0.40 &   -      & -&  -  & -  &  &  5\\
         &        &       &      &         &          &   &    &    &   &   \\
\hline
         &        &       &      &         &          &   &    &    &   &   \\
HD~100764 & 1.66$\pm$0.02  &  - &  1.08        &   -     &    -     &  -     &      -    &   -      & - &1\\
         &        &       &      &         &          &   &    &    &   &   \\
\hline
         &        &       &      &         &          &   &    &    &   &   \\
HD~198269 &  1.66$\pm$0.15 & 1.49$\pm$0.09 & 1.48$\pm$0.11 & 1.88$\pm$0.13&   -&   1.55 & 2.88 &  1.67&  2.4           & 1\\
         & 1.5$\pm$0.3    &    -          &1.2$\pm$0.1    &   1.2$\pm$0.2& -  & -      &   -  &   - &  2.4$\pm$0.2 &  2\\
         & 1.6$\pm$0.3    &   -           & 1.0$\pm$0.2   & 0.9$\pm$0.2  &  - &-       & -    & -  &  -            &  3\\
         &        &       &      &         &          &   &    &    &   &   \\
\hline
         &        &       &      &         &          &   &    &    &   &   \\
HD~224959 & 2.28$\pm$0.17 & 2.35$\pm$0.11 & 2.30$\pm$0.09 & 2.07$\pm$0.25 & 2.01&   -- & -&    - & 2.7      & 1\\
         & 1.9$\pm$0.3   &    -          &2.0$\pm$0.2    &   1.9$\pm$0.2  & -  &      - & - &   - &  3.1$\pm$0.2 &  2\\
         & 2.1$\pm$0.1   &   -           & 1.8$\pm$0.3   & 1.4$\pm$0.15    &  - &-      & - & -   &  -            &  3\\
         &        &       &      &         &          &   &    &    &   &     \\
\hline
\end{tabular}
\\
1. This work, 2. Van Eck et al. (2003),  3. Vanture (1992c), 5. Luck and 
Bond (1982)
\end{table*}
}

{\footnotesize
\centering
\begin{table*}

{\bf {Table 12 : Component coefficients and reduced $\chi^{2}_{\nu}$ from   
parametric model based analysis}}
\begin{tabular}{ l c  c c c c  }
\hline
         &            &          &          &                     \\
Objects  &   [Fe/H]   & A$_{s}$  & A$_{r}$  &  ${\chi^{2}_{\nu}}$ \\
         &            &          &          &                     \\
\hline
         &            &          &          &                      \\
HD~26    &  $-$1.13   & 0.69$\pm$0.09       & 0.31$\pm$0.08   & 3.2  \\
HD~198269&  $-$2.04   & 0.07$\pm$0.06       & 0.88$\pm$0.06   & 4.1   \\
HD~224959&  $-$2.42   & $-$0.03$\pm$0.08    & 0.98$\pm$0.07   &  5.4   \\
         &            &          &          &                     \\
\hline
\end{tabular}
\\
\end{table*}
}

\clearpage

\clearpage
\begin{figure*}
\epsfxsize=7truecm
%\epsffile{plotHD26simB.eps}
\epsffile{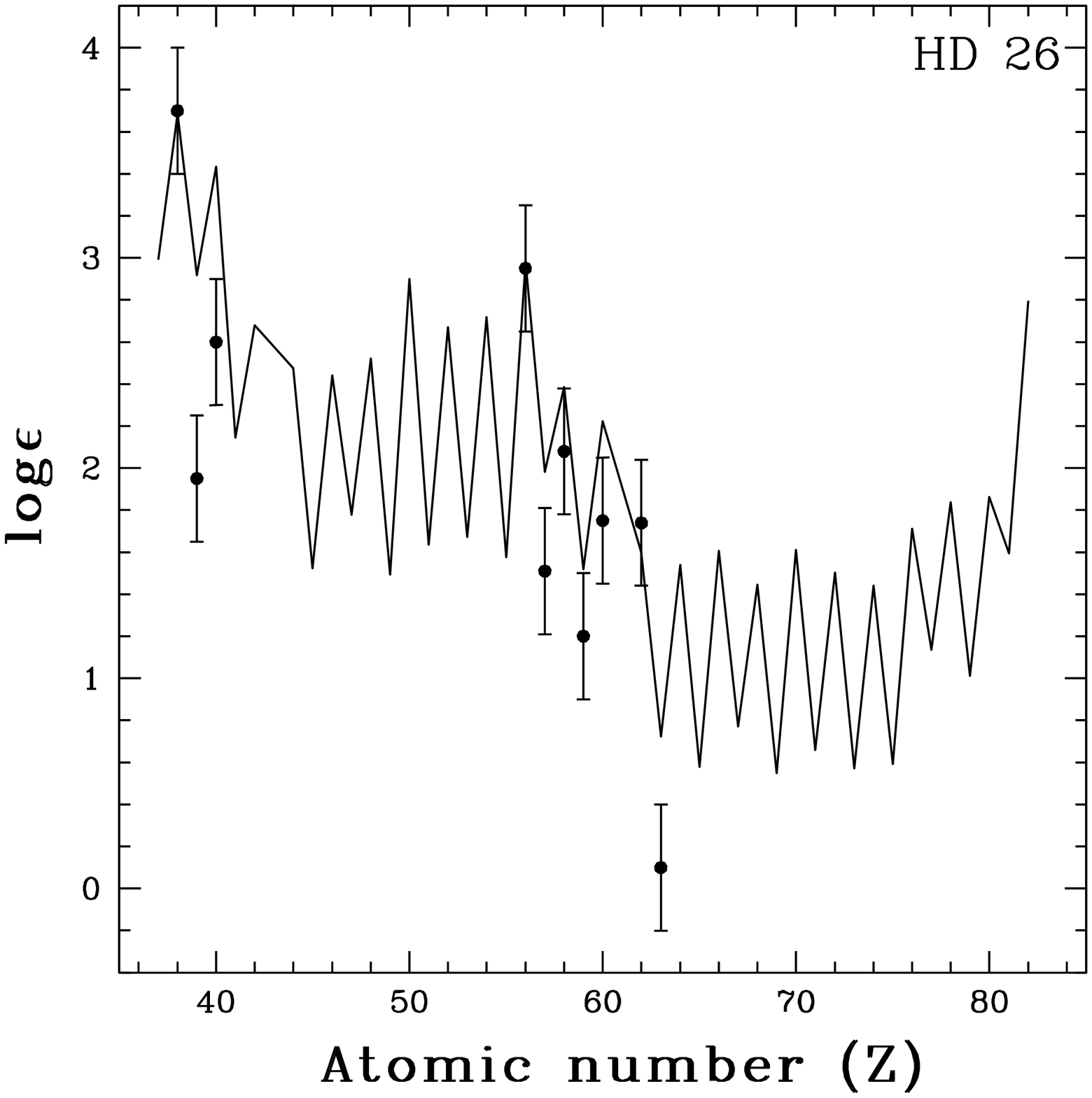}
\end{figure*}
\begin{figure*}
\epsfxsize=7truecm
\epsffile{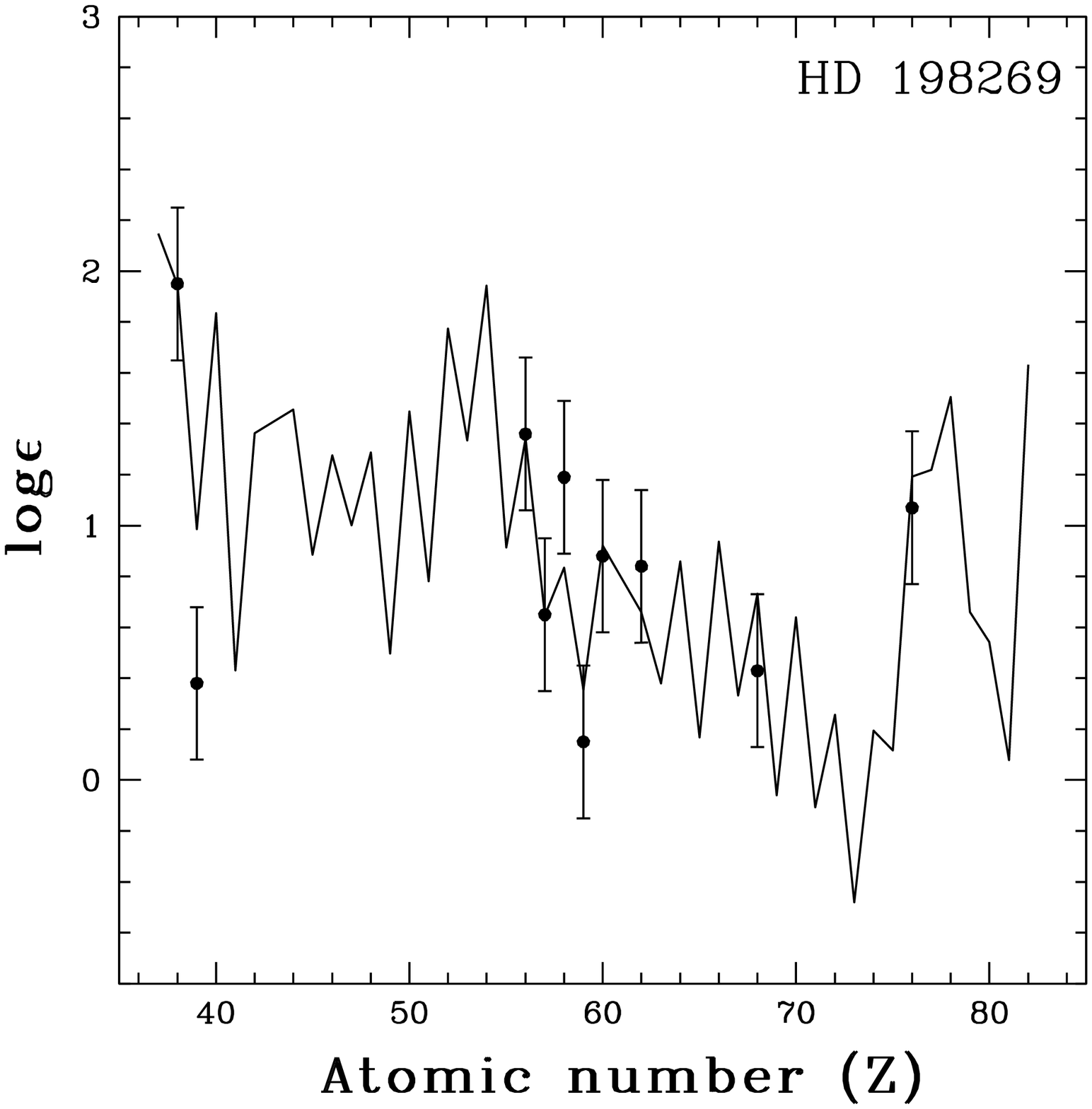}
\end{figure*}
\begin{figure*}
\epsfxsize=7truecm
%\epsffile{plotHD224959simB.eps}
\epsffile{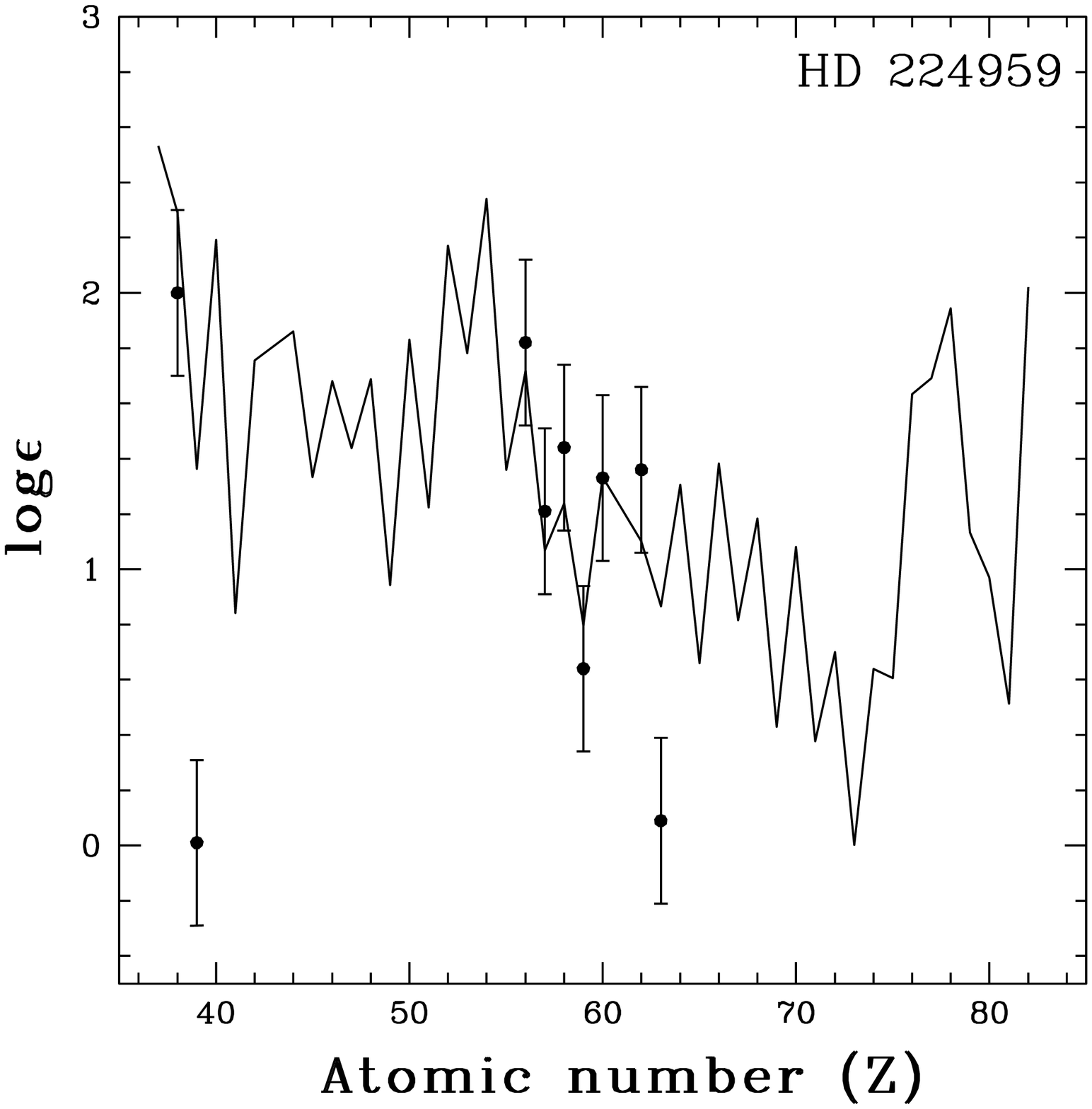}
\caption{Solid curves represent the best fit for the parametric model
function log${\epsilon}$  = A$_{s}$N$_{si}$ + A$_{r}$ N$_{ri}$, where N$_{si}$
and N$_{ri}$ represent the abundances due to s- and r-process respectively 
(Arlandini et al. 1999, Stellar model, scaled to the metallicity of the 
star). The best fit coefficients A$_{s}$, A$_{r}$ and  their respective
reduced chisquare are listed in Table 11.  The points with errorbars
indicate the observed abundances.}
%\label{Figure 7}
\end{figure*}

\clearpage
\begin{figure*}
\epsfxsize=7truecm
%\epsffile{plotHD26normB.eps}
\epsffile{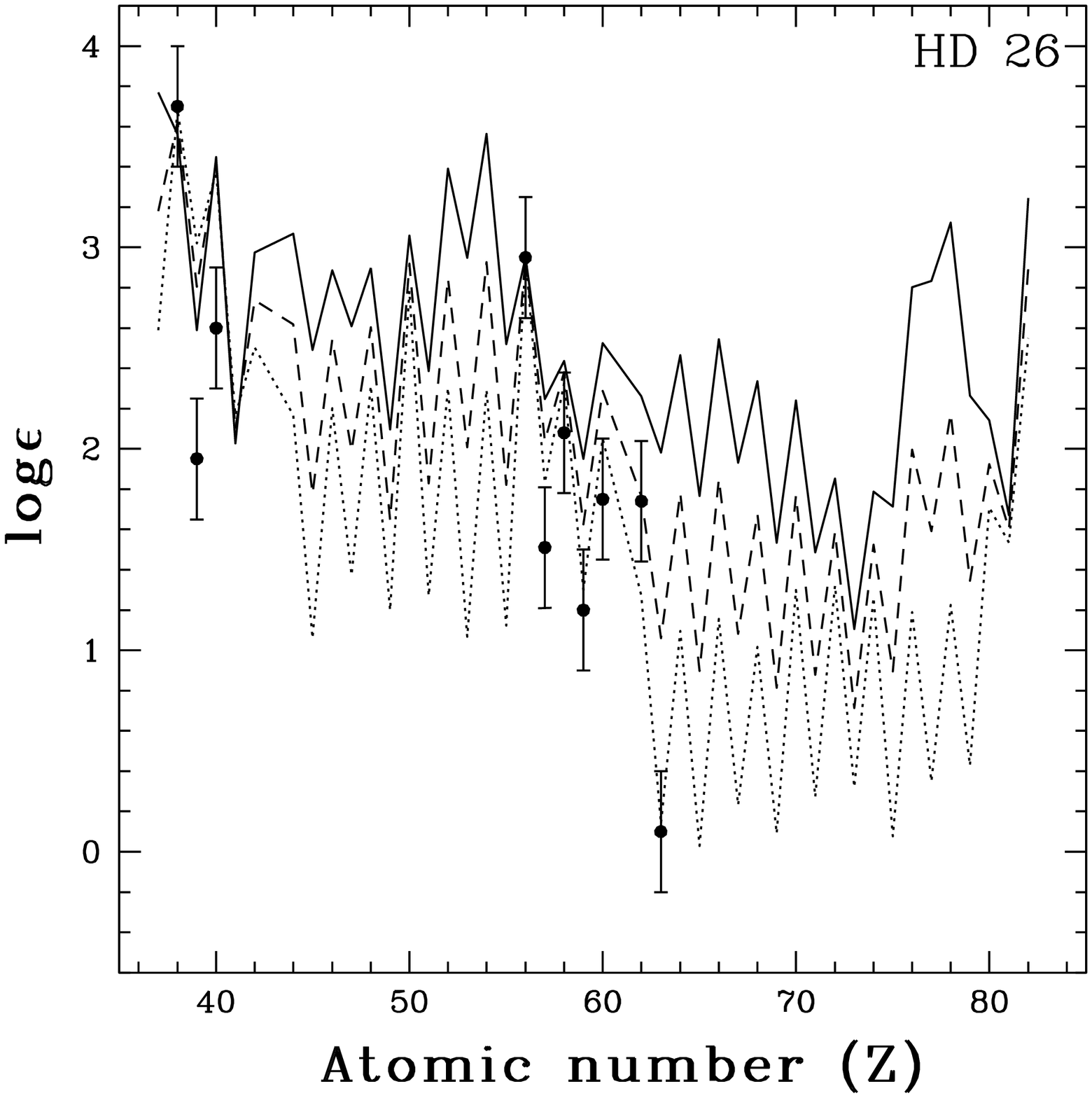}
\end{figure*}
\begin{figure*}
\epsfxsize=7truecm
%\epsffile{plotHD198269normB.eps}
\epsffile{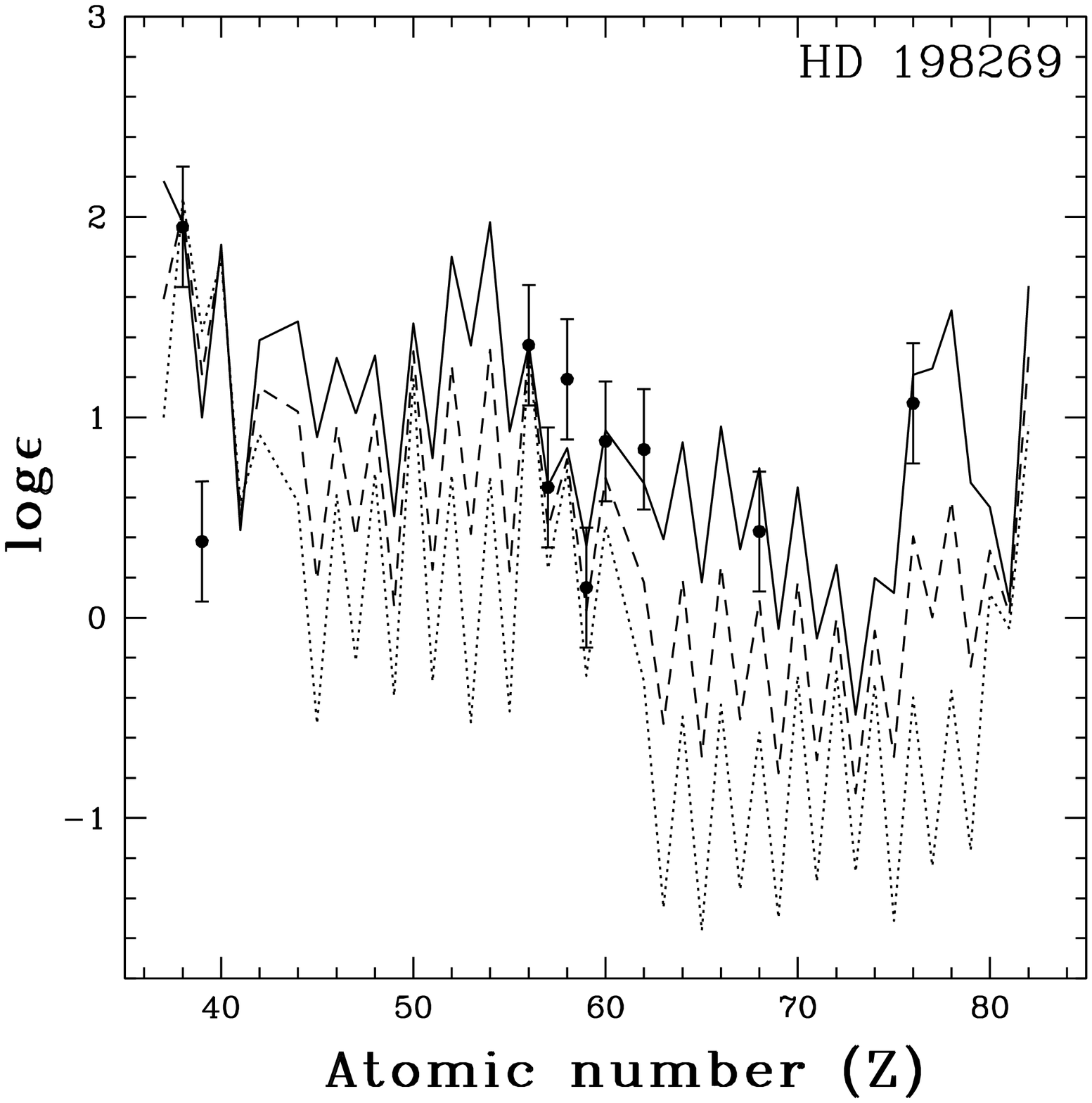}
\end{figure*}
\begin{figure*}
\epsfxsize=7truecm
%\epsffile{plotHD224959normB.eps}
\epsffile{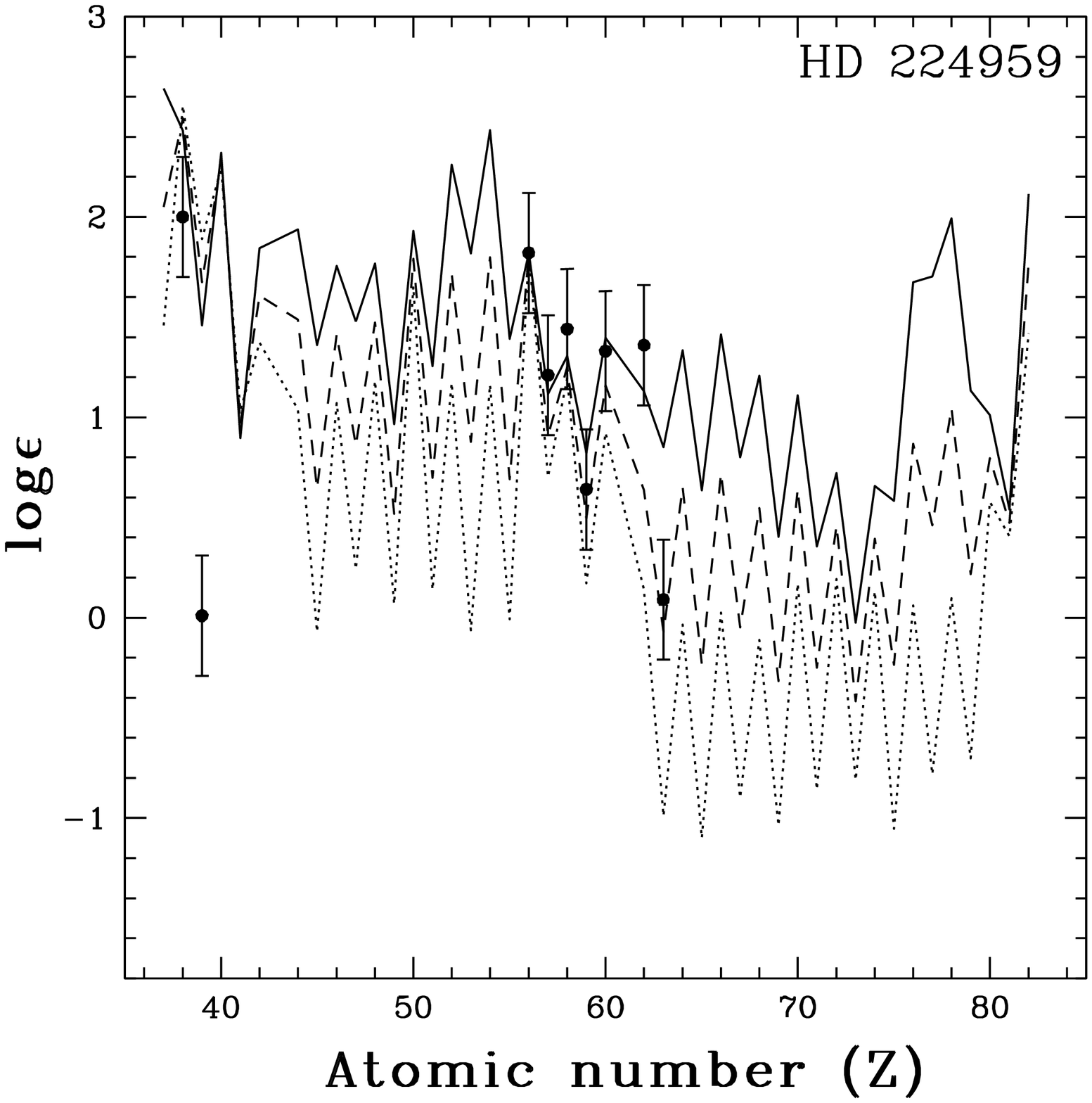}
\caption{
 Abundance  patterns  of heavy   elements from Arlandini et al. (1999)
(Stellar model). Solid line shows abundances due to only r-process;
dotted line is of s-process only and  dashed line indicates abundance
pattern derived from a simple average of  r- and s- processes. The abundances
 are scaled to the metallicity of the star and normalized to the observed Ba
abundances. The points with errorbars indicate the observed abundances.
}
%\label{Figure 8}
\end{figure*}
\clearpage

Calculated coefficients  indicate  dominance of r-process on abundances 
of  the heavy elements in case of HD~198269 and HD~224959 and the 
dominance of s-process in case of HD~26. It is noticed in  Figure 8, 
that the abundance pattern of elements  56 ${\le}$ Z ${\le}$ 63 in 
HD~26 agree with the s-process  pattern much better than with the r-process.

\section{Discussion and concluding remarks }

Results from abundance analysis  for three early type CH stars
and a carbon star with a cool dusty envelope  listed in the CH 
star catalogue (Bartkevicious 1996) are presented. 
The earlier chemical abundance studies of these objects were limited  to
 few elements and carbon isotopic ratios and  also limited
by both spectral resolution and wavelength region. We have conducted a 
detailed chemical composition  study of these objects using 
high-resolution Subaru spectra covering a 
wavelength  region from 4020 to 6775 \AA.
 We could achieve 
better accuracy in  abundance estimates with a  much lower
error range 0.1 - 0.3 dex compared to those reported previously
(i.e. Vanture 1992c). New abundance estimates for several neutron-capture
elements such as Sr, Y, Zr, Ba, La, Sm, Eu, Er, W, and Pb are
presented.  All the stars display larger enhancement of the 2nd-peak 
s-process elements such as  Ba, La, Ce, Nd, Sm etc.  compared to the 
1st-peak s-process elements Sr, Y, and Zr.  Compared to  Sun 
([Sr/Ba] = 0.74),  these stars  show a much smaller ratio of [Sr/Ba], 
i.e.  $-$0.04 (HD~26), $-$0.18 (HD~198269),  and $-$0.57 (HD~224959). In 
case of  HD~100764  we could measure abundances for only four heavy 
elements Y, La, Ce and Nd; and the  abundance ratios   
with [Y/Fe] = $-$0.08, [La/Fe] = 0.83, [Ce/Fe] = 1.66 and [Nd/Fe] = 1.08
are smaller compared to solar ratios. Abundance of Zn measured from 
a single  Zn I line gives a near-solar value for this object. While the 
abundances of Ba  could be estimated  for all the four objects,  the  
abundance estimate for Eu was possible only for HD~26 and HD~224959 
using our spectra. For HD~224959 our estimated  [Ba/Eu] = 0.06,  but  
HD~26 shows a high value with  [Ba/Eu]$\sim$ 1.32. 

Taking into consideration the range of [Ba/Fe], [Eu/Fe]
and [Ba/Eu]  (Beers \& Christlieb 2005, Masseron et al. 
2010, Jonsell et al. 2006) for  distinguishing CEMP-s stars from CEMP-r/s
stars,  we find that none of our program stars qualify to be a CEMP-r/s star.
At the non-availability of Eu abundance, CEMP stars with [Ba/Fe] ${\geq}$ 2.1
are classified as CEMP-r/s star (Masseron et al. 2010). None of our objects
show enhancement of Ba abundance as high as this value. HD~224959 with 
[Ba/Fe] = 2.07 is found to  marginally  meets  this criterion.
 HD~198269 with [Ba/Fe] = 1.24 also does not qualify to be a CEMP-r/s star, 
and same is the case for HD~26 with [Ba/Fe] = 1.93, [Eu/Fe] = 0.61 and 
[Ba/Eu] $>$ 1.32.  In the case of  HD~224959 the estimated [Eu/Fe] = 2.01  
satisfies the condition for  CEMP-r star (CEMP-r: [C/Fe] $>$ 1.0 and  
[Eu/Fe] $>$ +1.0); whereas with [Ba/Eu] = 0.06 this object also satisfies 
the condition for CEMP-r/s star  (CEMP-r/s: [C/Fe] $>$ 1.0 and  
0 $<$ [Ba/Fe] $<$ 0.5). As none of the  previous authors  reported on 
Eu abundance for these objects, a direct  comparison of Eu abundances 
estimated for HD~224959 and HD~26 was not possible.

 There are a lot of ambiguities in the classification of
lead stars,  as well as CEMP stars, in particular distinguishing between
 CEMP-s and CEMP-r/s stars. For  a well founded classification of 
these objects more tighter relations are needed.
The estimates of abundance ratios  of neutron-capture elements 
observed in the present sample  (Table 11), if
 extended for a larger  sample of CH stars covering a wide range in
metallicity, will allow  to set a limit on metallicity of stars 
showing enhanced abundances of s-process and/or both s- and r-process elements.

Except for   HD~100764, we have estimated the  abundance of lead, 
the 3rd-peak s-process element,  for the other three  stars in our 
sample. Pb  shows  large  enhancements  with [Pb/Fe] = 2.11 (HD~26), 
2.4 (HD~198269) and 3.7 (HD~224959). Enhancement of Pb abundance is 
noticed in a number of  CH as well as  CEMP-(r+s) stars. Van Eck 
et al. (2003)  defined stars with [Pb/hs] ${\ge}$ +1.0 as `lead  stars', 
where hs includes Ba, La and  Ce.  According  to  Jonsell et al. (2006) 
`lead stars' are those with [Pb/Ba] ${\ge}$ 1.0. This  definition 
does not require the `lead stars'  also to  be (r+s) stars, but 
requires that they are enhanced in Ba and Pb, and much more in the 
latter element. Estimated [Pb/Ba] in HD~26, 198269 and 224959 are 
respectively 0.74, 1.14 and 0.69. Following the  definition of Jonsell 
et al. 2006  the star HD~198269   satisfies the criteria to be a lead star. 
LP 625$-$44 and LP 706$-$7, both enriched in s-process elements
including Pb, were not considered as `lead stars' by Aoki et al. (2001)
as they show low values of [Pb/Ce] ${\leq}$ 0.4; alternatively, they suggest,
[Pb/Ce] ${\geq}$ 0.4 could be a criterion  for classifying a star 
as a `lead star'. This requirement in the abundance ratio of [Pb/Ce]
 is met by  all the three stars studied here with 
[Pb/Ce] = 0.46 (HD~26), 0.74 (HD~198269), 1.42 (HD~224959).

Models of very low-metallicity AGB stars developed in the framework of 
the partial mixing of protons into the deep carbon-rich layers predict
overabundance of Pb-Bi as compared to lighter s-elements. These stars are
characterized by large [Pb/Fe], [s/Fe] and [Pb/s] abundance ratios. In
this scenario, AGB stars with  [Fe/H] $\ge$ $-$1.3, are predicted to exhibit
[Pb/hs] (where hs refers to heavy s-process elements, such as Ba, La or Ce)
ratios as large as 1.5. These model predictions are  robust with respect to 
the model parameters, i.e., the abundance profile of the protons in the 
partially mixed layers or the extent of the partial mixing zone and 
uncertainties in the reaction rates (Van Eck et al. 2001). In agreement 
with the model predictions
the three CH stars under this study are found to exhibit overabundance
of Pb with respect to other s-process elements. High spectral resolution 
is essential to resolve the 4057.81 \AA\, Pb I line from the nearby CH line
at 4057.7 \AA\, and therefore to derive correct lead abundances.  We 
note that our spectra  do not have very good signal at this spectral 
region. Other Pb I lines
at 4063 \AA\, are not detectable in our spectra and Pb I 7228.97 \AA\, line is out of our wavelength region. 

Some of the previous reports (Sneden et al. 1998, 2000; Aoki et al. 2000)  
with large lead overabundances in low-metallicity stars are found not to 
 satisfy all these conditions. In these studies, the overabundances of lead
are attributed to the r-process rather than the s-process, because they
occur with strong enhancement of the r-process element Eu. For instance,
LP 625-44 shows [Pb/Fe] = 2.65, but [Pb/La] = 0.15 and [Pb/Ba] = $-$0.09
(Aoki et al. 2000).  Our estimated [Pb/La], [Pb/Ba], [Pb/Ce], [Pb/Sm] and [Pb/Nd] values are respectively (0.58, 0.18, 0.46, 0.23, 0.66) for HD~26,
(0.83, 1.16, 0.74, 0.52, 0.92) for HD~198269 and (1.2, 1.63, 1.42, 1.63, 1.4)
for HD~224959. Usually, [Pb/hs] is expected to decrease with  increasing 
metallicity from [Fe/H] = $-$2 to $-$1 (K\"appeler et al. 2011); our results 
are in agreement with this.

We have compared the derived abundances (scaled to Ba abundance) with the 
predicted elemental abundances at the surface of AGB stars at four different
metallicities (z = 0.018, z = 0.08, z = 0.004 and z = 0.001) in the dredge-up
material, 10th, 30th and 50th pulses of Goriely and Mowlavi (2000). In 
this scenario the primary $^{13}$C pocket formed in the proton-mixing zone
at low proton-to-$^{12}$C ratios is found to be responsible for an efficient
production of s-process nuclei. None of the model predictions show a 
proper match with the observed abundances.

In  Figure 8, we have compared the resulting abundances of neutron-capture 
elements for HD~ 26, 198269 and 224959  along with the scaled abundance 
patterns of the solar system material, the main s-process component, and  
the r-process pattern (Arlandini et al. 1999). The abundance patterns of 
elements with 56 ${\le}$ Z ${\le}$ 63  agree with the s -process pattern 
much better than with the r-process pattern indicating  that the 
neutron-capture elements in HD~26 principally originate in the s-process. 
Estimated [Ba/Eu] (= 1.32) for HD~26 is similar to that  seen in the 
main s-process component ([Ba/Eu] = 1.15, Arlandini et al. 1999), but 
much lower in the case of HD~224959 with  [Ba/Eu] = 0.06. The abundance 
patters of elements 56 ${\le}$ Z ${\le}$ 63 are however in agreement 
with the s-process pattern.  The observed low values of [Ba/Eu]  in 
HD~224959 may be  interpreted  as a result of an s-process that produces
different abundance ratios from that of the main s-process component
(Aoki et al. 2002).  This idea needs further investigation. 

Theoretical models  of  Goriely \& Mowlavi (2000)  predicted [Ba/Eu] = 0.4 
for yields of metal-deficient AGB stars. Analyses of the observed 
abundances of the heavy elements for HD~26 based on a  parametric  
model  show that s-process have  higher  contributions  with    
$A_s$  ${\sim}$ 0.69 than  r-process with  $A_r$ ${\sim}$  0.31, where   
$A_s$ and $A_r$ are the component coefficients that correspond to 
contributions  from  the s- and r-process  respectively. Here,  the 
contributions from   r- and s-process  are estimated from the fitting 
for Ba - Er. 

High velocities are a common feature of CH stars.  Temporal variations 
of radial velocities observed among the known CH stars indicate 
binarity of the objects. This is in support of the  widely accepted 
scenario for the formation of CH  stars, involving   mass transfer from a
companion AGB star  (McClure 1983, 1984 and McClure and Woodsworth 1990).
While the radial velocity estimate for HD~100764 is $\sim$ 5 km s$^{-1}$,
the other three  objects are high-velocity stars (Table 2). The high radial
velocities suggest halo membership.  Our estimates  of radial velocities 
compare well with  literature estimates that confirmed these objects to 
be  radial velocity variables  (McClure and Woodsworth  1990) with  
periods of 3.55 yrs (HD~198269)   and 3.49 yrs (HD~224959). Kinematic 
properties support a physical scenario in which the observed enhancement
of neutron-capture elements  resulted from a transfer of material rich 
in s-process elements across a binary system with an AGB star.

{\it Acknowledgment}\\
 This work made use of the SIMBAD
astronomical database, operated at CDS, Strasbourg, France, and the NASA
ADS, USA. Funding from the DST project SB/S2/HEP-010/2013  
 is gratefully acknowledged.
\\

\newpage

{\footnotesize
\begin{table*}
\centering
{\bf Table 13: Equivalent widths}\\
\begin{tabular}{|l|l|c|c|r|c|c|c|c|c|}
\hline
          &      &       &             &               &       &              &              &             &    \\
Wlab      & id   &   Z  &      Elow   &        log gf  & HD~26  &     HD~100764 &     HD~198269 &   HD~224959 &  Ref    \\ 
          &       &      &             &               &       &              &              &            &     \\
\hline
  5889.95& Na I  &   11.0   &    0.00     &    0.10    &  324.3 &   -      &   258.5  &    184.9& 1    \\
  5895.92&    &   11.0   &    0.00     &   -0.20    &  284.3 &   -      &   240.4  &    175.6 & 1     \\
  4057.50& Mg I   &   12.0   &    4.35     &   -0.89    &    -   &   -      &   98.35  &      -   &2    \\
  4571.10&    &   12.0   &    0.00     &   -5.69    &    -   &   -      &   151.9  &    57.76 &2    \\
  5172.69:&   &   12.0   &    2.71     &   -0.38    &    -   &   -      &   324.7  &    185.8 &2    \\
  5528.40 &   &   12.0   &    4.35     &   -0.34    & 184.6  &   -      &   151.0  &    122.6 &3    \\
  5588.76&Ca I    &   20.0   &    2.53     &    0.36    &    -   &   -      &   121.0  &     57.3&3    \\
  5594.47&    &   20.0   &    2.52     &    0.10    &    -   &   -      &   118.8  &     -  &3      \\
  5598.49&    &   20.0   &    2.52     &   -0.09    &    -   &   -      &   116.9  &     -   &3     \\
  5857.45&    &   20.0   &    2.93     &    0.24    &   -    &   -      &   95.19  &     -   &3     \\
  6102.72&    &   20.0   &    1.88     &   -0.77    & 117.2  &   -      &   110.0  &   41.7  &3    \\
  6122.22&    &   20.0   &    1.89     &   -0.32    & 151.4  &   -      &   143.8  &    -    &3     \\
  6162.17 &   &   20.0   &    1.90     &   -0.09    &    -   &   -      &   164.4  &   93.6  &3    \\
  6439.07 &   &   20.0   &    2.53     &    0.39    & 144.4  &   -      &   137.0  &   73.4 &3     \\
  6462.567:&  &   20.0   &    2.52     &    0.26    &    -   &   -      &     -    &  182.2 &3      \\
  4415.54  & Sc II &   21.1   &    0.60     &   -0.89    &    -   &   -      &   88.56  &    -   &3      \\
  5031.01 &   &   21.1   &    1.36     &   -0.40    &    -   &   -      &   107.6  &     -  &3      \\
 $^\#$6245.630&  &   21.1& 1.51       &  -1.03   & hfs-syn    &  --     &   hfs-syn    &  hfs-syn &  3  \\
  4533.239&Ti I   &   22.0   &    0.85     &    0.48    & 113.0  &   -      &   105.7  &    54.2 &4    \\
  4534.778&   &   22.0   &    0.84     &    0.28    & 108.1  &   -      &   105.6  &    60.2 &4    \\
  4981.73 &   &   22.0   &    0.85     &    0.50    &    -   &  -       &   124.7  &    -    &4    \\
  4991.067&   &   22.0   &    0.84     &    0.38    &    -  &    -      &   126.4  &    57.6&4     \\
  5039.96 &   &   22.0   &    0.02     &   -1.13    &    -   &   -      &   107.8  &      -  &4     \\
  5064.65 &   &   22.0   &    0.05     &   -0.99    &    -   &  -       &   115.2  &     -  &4     \\
  5210.386&   &   22.0   &    0.05     &   -0.88    &    -  &   -       &   117.6  &    49.2 &4    \\
  4025.12:&Ti II   &   22.1   &    0.61     &   -1.98    &    -  &   -       &   119.9  &    -    &4     \\
  4053.83:&   &   22.1   &    1.89     &   -1.21    &    -  &   -       &   67.69  &    -    &4     \\  
  4287.88:&   &   22.1   &    1.08     &   -2.02    &    -  &    -      &   122.3  &   52.4  &4    \\
  4344.30:&   &   22.1   &    1.08     &   -2.09    &    -  &    -      &   126.0  &     -   &4     \\
  4418.31&    &   22.1   &    1.24     &   -1.99    &  120.0&     -     &   72.43  &    -    &4      \\  
  4443.77 &   &   22.1   &    1.08     &   -0.70    &163.0   &  -       &   168.4  &   113.7  &4    \\
  4450.50&    &   22.1   &    1.08     &   -1.51    &    -   &   -      &   135.5  &   79.9  &3     \\
  4464.46: &  &   22.1   &    1.16     &   -2.08    &    -   &  -       &   116.1  &    -    &4     \\
  4501.27&    &   22.1   &    1.12     &   -0.76    &    -   &  -       &   164.7  &   122.5 &4     \\
  4529.48:&   &   22.1   &    1.57     &   -2.03    &    -   &   -      &   101.7  &    -    &4     \\
  4563.77 &   &   22.1   &    1.22     &   -0.96    &173.1   &  -       &   160.2  &   119.4 &4     \\
  4571.96 &   &   22.1   &    1.57     &   -0.53    &     -  &  -       &   187.1  &    -    &4     \\
  4589.92:&   &   22.1   &    1.24     &   -1.79    &    -   &  -       &   107.1  &    57.0 &4    \\
  4657.21 &   &   22.1   &    1.24     &   -2.32    &   -   &    -      &   74.6  &    -     &4    \\ 
  4779.98 &   &   22.1   &    2.05     &   -1.37    &    -  &    -      &   74.0  &    -     &4    \\
  4798.51 &   &   22.1   &    1.08     &   -2.67    &    -  &   -       &   86.2  &    -   &3      \\
  4805.09 &   &   22.1   &    2.06     &   -1.10    &    -   & -        &   105.4  &    68.4 &4    \\
  4865.61 &   &   22.1   &    1.12     &   -2.81    & 77.1  &  -        &   65.3  &    -     &4    \\
  5185.90 &   &   22.1   &    1.89     &   -1.35    &    -   &  -       &   87.0  &   38.3  &4    \\
  5226.53 &   &   22.1   &    1.57     &   -1.30    &     -  &  -       &   122.9  &     -  &4      \\
 $^\#$ 5727.048 &V I&  23.1   &  1.08      & -0.012     & hfs-syn    &  --     &  hfs-syn    &  hfs-syn &  3  \\ 
  4254.33 &Cr I  &   24.0   &    0.00     &   -0.11    &    -   &  -       &   183.9  &      - &4     \\
  4600.75 &   &   24.0   &    1.00     &   -1.26    &    -   & -        &   79.4   &      - &4     \\
  4616.12 &   &   24.0   &    0.98     &   -1.19    &    -   & -        &   99.0   &      - &4     \\
  4626.19 &   &   24.0   &    0.97     &   -1.32   &  88.2  &  -       &    -     &     -   &4   \\
  4646.17&    &   24.0   &    1.03     &   -0.70    &     -  &  -       &   102.9  &     41.8 &4 \\
  4652.16&    &   24.0   &    1.00     &   -1.03    &     -  & -        &   107.0  &     -   &4   \\
  5206.04 &   &   24.0   &    0.94     &    0.02    &    - &  -        &   162.4  &    78.7  &4   \\
  4451.58:&Mn I  &   25.0   &    2.89     &    0.28    &    - &  -        &   118.0  &     -    &4  \\
  4754.05 &   &   25.0   &    2.28     &   -0.09    &    - &  -        &   77.3   &      -   &4  \\
  4783.43 &   &   25.0   &    2.30     &    0.04    &    - &  -        &   100.7  &       -  &4   \\ 
  4823.53&    &   25.0   &    2.32     &    0.14    &    - &  -        &   106.5  &      -    &4 \\
$^\#$   6013.488 &  &   25.0      &  3.072 & -0.250   & syn    &  syn      &  syn    &  syn &  4\\ 
\hline 
\hline
\end{tabular}
\end{table*}
}

\clearpage
{\footnotesize
\begin{table*}
{\bf Table 13: Equivalent widths (continued)}\\
\begin{tabular}{|l|l|c|c|r|c|c|c|c|c|}
\hline
          &      &       &             &               &       &              &              &             &    \\
Wlab      &       id  &  Z&      Elow   &     log gf   &     HD~26  &     HD~100764 &     HD~198269 &   HD~224959&Ref   \\ 
          &      &       &             &               &       &              &              &             &    \\
\hline
  4855.41& Ni I    &   28.0   &    3.54     &    0.00   &   -  &  -        &   51.8   &    18.6 & 5   \\
  4980.16&     &   28.0   &    3.61     &   -0.11   & 89.5 &  -        &   53.8   &     - & 5      \\
  5035.37&     &   28.0   &    3.63     &    0.29   & 79.2 &  -        &   60.4   &    -  & 5    \\
  5080.52&     &   28.0   &    3.65     &    0.13   &    - &  -        &   65.0   &     - & 5     \\
  5137.08&     &   28.0   &    1.68     &   -1.99   &    - &  -        &   93.0  &    -  & 5     \\
  4810.53&     &   30.0   &    4.08     &   -0.17   & 84.2 &  -        &     -    &  -  &3       \\
  4607.327&Sr I    &   38.0   &    0.00     &   -0.57   & syn  &  syn      &  syn  &  syn &  6  \\
  6550.24:&    &   38.0   &    2.69     &   +0.18   &  -   &  104.8    &   42.6   &   -&  6      \\
  4883.685&Y II    &   39.1   &    1.08     &    0.07   &149.4 &  101.0    &   113.9  &    - & 7     \\
  5087.43 &    &   39.1   &    1.08     &   -0.17   &   -   &  229.9   &     -    &    - &  7    \\
  5200.413 &   &   39.1   &    0.99     &   -0.57   & syn &  syn  & syn  &  syn   &  7    \\
  5205.73 &    &   39.1   &    1.03     &   -0.34   &     - &  -       &   86.2   &   45.3&  7      \\
  5473.388:&    &   39.1   &    1.74     &   -1.02   &     - & -        &   23.6   &  -  &  7        \\
  4414.124:& Zr I   &   40.0   &    0.15     &   -2.32   &     - & 56.09    &   67.2   &  -  &  6        \\
  5311.434:&   &   40.0   &    0.52     &   -1.71   &     - &  -       &   54.3   &  -    &  6      \\
  6134.585:&   &   40.0   &    0.00     &   -1.28   &    syn & -   &   syn    & 
  syn    &  8   \\
  4443.00: & Zr II &   40.1   &    1.49     &   -0.33   &     - & -        &  254.3   &  -  &  8      \\
  4496.97:&     &   40.1   &    0.71     &   -0.59   &     - & -        &  20.6    &  - &  8      \\
  4812.281 &Mo I   &   42.0   &    3.26   &   -1.90   &     -  &  -          &  34.4    &  - &9        \\
  5204.552 &   &   42.0   &    3.36   &   -1.87   &      - & 253.5       &  37.2    &  100.8 &9     \\
  5298.010&    &   42.0   &    2.68   &   -1.85   &     -  &  -          &  207.3   &    -  &9      \\
  5991.370 &   &   42.0   &    3.44   &   -1.64   &     -  &  73.7       &  25.7    &    -  &9      \\
  5752.022& Rb I   &   44.0   &    1.00   &   -2.96   &    -   &  -          &  18.2    &     - &10     \\
  4554.036& Ba II   &   56.1   &    0.00   &   0.12    &     -  & 250.7       &      -   &     -  &2    \\
  4934.076&    &   56.1   &    0.00   &   -0.15   &     -  &  292.7      &  356.5   &   332.1 &1   \\
$^\#$  5853.668&    &   56.1   &    0.60   &   -1.02   &    hfs-syn   & 153.2       &   hfs-syn   &  hfs-syn &1   \\
  6141.727&    &   56.1   &    0.70   &   -0.08   &    syn &  syn       &   250.8  &   245.5  &1  \\
  6496.897&    &   56.1   &    0.60   &   -0.38   &    -   & 235.7       &   252.1  &   -    &1    \\
  4086.709& La II   &   57.1   &    0.00   &   -0.15   &    -   &  116.2      &     -    &     -  &11   \\
  4322.51: &    &   57.1   &    0.17   &   -1.05   &   -    &  -          &   102.1  &  97.4  &12    \\
  4526.111:&    &   57.1   &    0.77   &   -0.77   & 81.9   &    46.4     &   57.7   &    - &11     \\
  4580.045 &   &   57.1   &    0.71   &    -1.02  &  -     &     -       &     -    &  51.5 &12    \\
  4662.51: &    &   57.1   &    0.00   &   -1.28   & 99.0   &    -        &   93.2   &    - &12      \\
  4748.726 &   &   57.1   &    0.93   &   -0.86   & 71.2   &    -        &   48.7   &  51.3 &12     \\
  4840.00:&    &   57.1   &    0.32   &   -2.07   &    -   & 9.5         &   44.2   &  44.8 &12      \\
$^\#$4921.776&   &      &    0.24  &  -0.68   & hfs-syn &  --      &  hfs-syn  &  hfs-syn  &12  \\
  4999.461 &  &   57.1   &    0.40   &   -0.89   &    -   & 144.2       &      -   &    -  &12      \\
  5259.38: &  &   57.1   &    0.17   &   -1.76   &     -  &   -         &   43.9   &   -   &12      \\
  6320.43: &   &   57.1   &    0.17   &   -1.52   &     -  &   -         &   74.0   &   -  &13      \\
  5328.085: &Ce I  &   58.0   &    0.49   &    0.01   &     -  &   -         &   225.7  &   - & 14      \\
  6093.193 &  &   58.0   &    0.80   &   -0.37   &     -  & 108.1       &   -   &   -    & 14    \\
  6335.340 &  &   58.0   &    0.47   &   -1.32   &     -  & 133.6       &   -   &   -    & 14    \\
  4062.22 &Ce II   &   58.1   &    1.37   &    0.26   &     -  &   -         &   41.9   &  36.0 & 14     \\
  4076.24 &   &   58.1   &    0.81   &   -0.42   &     -  &  -          &   43.7   &  -   & 14      \\ 
  4117.29 &   &   58.1   &    0.74   &   -0.525   &     -  &  -          &   42.4   &  36.0 & 14     \\
  4127.38:&    &   58.1   &    0.68   &   +0.11   &     -  &  -          &   69.4   &  53.1& 14      \\
  4185.33: &   &   58.1   &    0.42   &   -0.64   &     -  &  -          &   48.7   &  -   & 14     \\
  4190.63  &  &   58.1   &    0.90   &   -0.47   &     -  &  -          &   180.8  &    -  & 14    \\
  4222.60: &   &   58.1   &    0.12   &   -0.30   &     -  &  -          &   97.7   &  80.2& 14    \\
  4257.12  &  &   58.1   &    0.46   &   -1.12   &     -  & -           &   49.6   &   -   & 14    \\
  4336.244 &  &   58.1   &    0.70   &   -0.56   & 84.4   &  -          &   62.7   &     - & 14    \\
  4423.673 &  &   58.1   &    1.06   &   -0.32   &     -  &  -          &   54.3   &   46.6& 14   \\  
  4427.92  &  &   58.1   &    0.54   &   -0.46   &     - &  -       &   84.9     &    -    & 14 \\
  4486.91: &   &   58.1   &    0.30   &   -0.36   &    -  & 14.4     &   96.0     &   77.1 & 14  \\
  4515.86: &   &   58.1   &    1.06   &   -0.52   & 56.3  &     -    &   42.2     &   41.1 & 14  \\
  4522.079 &  &   58.1   &    0.87   &   -1.49   &   -   & 28.6     &   35.9     &       - & 14  \\
  4523.075 &  &   58.1   &    0.51   &   -0.30   &     - &  -       &   116.3    &       - & 14  \\
  4527.348 &  &   58.1   &    0.32   &   -0.46   &     - & 142.6    &     -      &       - & 14  \\
  4539.745 &  &   58.1   &    0.32   &   -0.45   &     - &  -       &   96.8     &    73.5& 14 \\
  4544.96  &  &   58.1   &    0.42   &   -0.97   &     - &  -       &   94.9     &      - & 14  \\ 
  4551.291: &  &   58.1   &    0.74   &   -0.58   &     - & 116      &   78.6     &       -& 14  \\
\hline 
\hline 
\end{tabular}
\end{table*}
}

{\footnotesize
\begin{table*}
{\bf Table 13 : Equivalent widths (continued)}\\
\begin{tabular}{|l|l|c|c|r|c|c|c|c|c|}
\hline
          &       &     &            &           &       &           &           &         &  \\
Wlab      &       id &Z  &      Elow  &    log gf  & HD~26 & HD~100764 & HD~198269 & HD~224959&Ref \\ 
          &          &  &            &           &       &           &           &          &  \\
\hline
  4562.37 &   &   58.1   &    0.48   &   +0.33   & 120.1 &  72.0    &     -      &   99.7   &3\\
  4628.16 &   &   58.1   &    0.52   &   +0.26   &     - &  -       &   108.0    &       - &14   \\
  4744.944 &  &   58.1   &    0.44   &   -1.67   &     - &   -      &   47.1     &    -   &14   \\
  4747.167 &  &   58.1   &    0.32   &   -1.24   &     - &  -       &   56.0     &      -  &14  \\
  4757.841: &  &   58.1   &    0.95   &   -0.78   &     - &  -       &   43.9     &     - &14    \\
  4773.941 &  &   58.1   &    0.92   &   -0.49   &    -  &          &   51.4     &   38.9  &14  \\
  4873.999:&   &   58.1   &    1.10   &   -0.89   &    -  & 50.7     &   44.9     &    28.4&14   \\
  5022.867:&   &   58.1   &    1.04   &   -0.72   &89.7   & 117.4    &   81.8     &   37.8 &14   \\
  5187.457:&   &   58.1   &    1.21   &   -0.10   &  -    &40.15     &   73.0     &   49.6 &14    \\
  5274.230 &  &   58.1   &    1.04   &    0.32   & 92.1  &  94.5    &   71.3     &   59.2  &14  \\
  5330.556:&   &   58.1   &    0.86   &   -0.76   & 76.7  & 23.89    &   55.9     &    38.3&14   \\
  5512.064 &  &   58.1   &    1.00   &    -0.51   &   -   &  -       &   53.5     &    34.1 &14   \\
  5975.818:&   &   58.1   &    1.32   &   -0.81   &   -   &  -       &   31.1     &    18.3 &14 \\
  5996.060 &Pr I  &   59.0   &    1.20   &   -0.17   &    -  & 37.7     &      -     &  -   &  15  \\
  4148.44  &Pr II  &   59.1   &    0.22   &   -0.72   &   -   &  -       &   81.1     &    -  &  15    \\ 
  5188.217 &  &   59.1   &    0.92   &   -1.14   &   -   &  -       &   24.2     &   -  &  15    \\
  5219.045 &  &   59.1   &    0.79   &   -0.24   &   -   &   -      &   34.4     &  -    &  16   \\
  5220.12  &  &   59.1   &    0.80   &   +0.17   &   -   &   -      &   58.5     &  58.2 &3  \\
  5259.74  &  &   59.1   &    0.63   &   -0.07   & 86.0  &   -      &   57.8     &   55.9  &  16 \\
  5292.619 &  &   59.1   &    0.64   &   -0.30   &69.4   &  52.5    &   32.6     &  30.6 &  16   \\
  5322.772:&   &   59.1   &    0.48   &   -0.31   &94.9   &     -    &   63.8     &   62.1 &  15   \\
  5331.460 &  &   59.1   &    0.64   &   -1.97   &26.5   &   60.8   &       -    &     - &  5   \\
  4051.141 &Nd II  &   60.1   &    0.38   &    0.09   &     - &   -      &   103.3    &      - &  17   \\
  4059.951 &  &   60.1   &    0.20   &   -0.36   &     - &  -       &   70.1     &    -  &  17    \\
  4061.08: &  &   60.1   &    0.47   &    0.55   & 130.7 &   -      &   115.8    &    -  &  15   \\ 
  4069.27: &  &   60.1   &    0.06   &   -0.40   & 101.6 &   -      &   92.2     &    -  &  17   \\
  4109.46  &  &   60.1   &    0.32   &   +0.28   &     - &   -      &   184.2    &    -   &  15   \\
  4133.36  & &   60.1   &    0.32   &   -0.34   &     - &   -      &   102.9    &     -  &  17  \\
  4232.38: &   &   60.1   &    0.06   &   -0.35   &     - &  -       &   115.5    &   91.5 &  18 \\
  4256.821 &  &   60.1   &    0.18   &   -1.39   &     - &  -       &   44.0     &   38.4 &  17   \\
  4412.256 &  &   60.1   &    0.06   &   -1.42  &    -  &   -     &   89.3      &     -  &  15   \\
  4446.39  &  &   60.1   &    0.20   &   -0.63  &     - &   -     &   102.3     &     -  &  15   \\
  4451.563: &  &   60.1   &    0.38   &   -0.04  &134.3  &131.5    &   118.0     &   108.0&  20   \\
  4451.98: &  &   60.1   &    0.00   &   -1.34  &   -   & 63.3    &   89.2      &   71.8  &  15   \\
  4462.979 &  &   60.1   &    0.55   &    0.07  &     - &  -      &   125.2     &      -  &  15   \\
  4594.445 &  &   60.1   &    0.20   &   -1.55  & 64.7  &    -    &   44.7      &    36.4 &  15   \\
  4749.560 &  &   60.1   &    0.55   &   -1.55  &  77.7 &   -     &   34.8      &     -   &  15   \\
  4797.153 &  &   60.1   &    0.55   &   -0.95  &     - &   -     &   49.2      &    46.5 &  15   \\
  4819.64: &  &   60.1   &    0.18   &   -2.26  & 46.9  &  18.4   &   42.2      &      -  &  15   \\
  4820.339: &  &   60.1   &    0.20   &   -1.24  &    -  &  -      &    -        &   68.0 &  15    \\
  4825.478 &  &   60.1   &    0.18   &   -0.86  &     - &  -      &   98.3      &      -  &  15  \\
  4859.039:&   &   60.1   &    0.32   &   -0.83  &111.3  &  -      &   89.2      &   70.3 &  15    \\
  4876.112:&   &   60.1   &    0.55   &   -1.67  &   -   &  -      &   34.8      &    30.3 &  15    \\
  4989.953:&   &   60.1   &    0.68   &   -0.50  &     - &  -      &   100.1     &    97.6 &  15   \\
  5200.121:&   &   60.1   &    0.55   &   -0.49  &     - &  -      &   69.4      &      - &  15     \\
  5212.361 &  &   60.1   &    0.20   &   -0.87  & 95.2  &  102.2  &   75.5      &    66.1 &  15   \\
  5234.21: &   &   60.1   &    0.55   &   -0.46  &   -   &    -    &   86.5      &       - &  18    \\
  5249.576:&   &   60.1   &    0.97   &    0.21  & 108.2 &  43.1   &   84.1      &    71.4 &  18   \\
  5255.506 &  &   60.1   &    0.20   &   -0.82  &    -  &   -     &   91.6      &      -   &  15   \\    
  5276.878 &  &   60.1   &    0.85   &   -0.44  & 77.3  &   -     &   49.4      &      -  &  18  \\
  5287.133 &  &   60.1   &    0.74   &   -1.30  & 39.8  &   -     &     -       &    19.7  &  15  \\
  5293.169 &  &   60.1   &    0.82   &   -0.06  &    -  &  91.9   &     -       &    75.9  &  18  \\
  5311.48 &   &   60.1   &    0.99   &   -0.42  &    -  &   -     &   54.3      &    53.8 &  18  \\
  5319.82 &  &   60.1   &    0.55   &   -0.21  & 118.3 &   -     &   97.1      &   89.8   &  18 \\
  5594.416: &  &   60.1   &    1.12   &   -0.04  &    -  &   -     &   118.8     &    73.1&  17  \\
  5688.52  &   &   60.1   &    0.99   &   -0.25  &   -   &   -     &   58.2     &     53.7 &  17 \\
  5718.118 &  &   60.1   &    1.41   &   -0.34  &    -  &   -     &   29.8      &     33.1&  17  \\
  5825.857 &  &   60.1   &    1.08   &   -0.76  &    -  &   -     &   30.2      &       - &  15   \\ 
  6367.403& Sm I  &   62.0   &    0.18   &   -1.86  &   -   &  59.4   &   20.6      &   31.1  &  19  \\
\hline 
\hline
\end{tabular}
\end{table*}
}

\clearpage
{\footnotesize
\begin{table*}
{\bf Table 13:  Equivalent widths (continued)}\\
\begin{tabular}{l|l|c|c|r|c|c|c|c|c|}
\hline
          &        &     &             &         &       &           &            &  &          \\
Wlab      &       id &  Z &      Elow   &  log gf  & HD~26 & HD~100764 &  HD~198269 &  HD~224959&Ref \\ 
          &          &   &             &         &       &          &             &         &   \\
\hline
  4220.66 & Sm II  &   62.1   &    0.54   &   -1.11  &   -   &   -     &   46.8      &    38.5 & 20 \\
  4244.70: &   &   62.1   &    0.28   &   -1.07  &   -   &   -     &   42.4      &    43.9&14   \\
  4318.94  &  &   62.1   &    0.28   &   -0.61  &   -   &   -     &   90.4      &    76.9 &14  \\
  4424.337 &  &   62.1   &    0.48   &   -0.26  &   -   &   -     &    91.9     &     -   &14   \\
  4434.318  & &   62.1   &    0.37   &   -0.57  & 100.1 &   -     &   62.5      &    - &14\\
  4458.509: &  &   62.1   &    0.10   &   -1.11  &    -  &   85.1  &   69.7      &    57.8 &14   \\
  4472.409: &  &   62.1   &    0.18   &   -1.36  &    -  &   -     &   38.3      &     -   &14   \\
  4499.475  & &   62.1   &    0.24   &   -1.41  &  62.3 &   -     &   50.1      &     41.4 &14  \\
  4519.63 &   &   62.1   &    0.54   &   -0.75  &   -   &         &   75.7      &    75.4 &14  \\
  4543.943 &  &   62.1   &   0.33  &   -1.00   &   -    &    169.8 &  116.2     &   80.0  &14 \\
  4566.21  &  &   62.1   &  0.33   &   -1.25   & 71.7   &     -    &   54.0     &    -   &14   \\   
  4577.69  &  &   62.1   &  0.25   &   -1.23   &   -    &     -    &   65.5     &    -   &14  \\
  4584.83 &   &   62.1   &  0.43   &   -1.08   &   -    &     -    &   64.5     &   75.2 &14  \\
  4593.54: &   &   62.1   &  0.38   &   -1.30   &   -    &    -     &   52.8     &     - &14  \\
  4595.283&   &   62.1   &  0.48   &   -1.02   & 105.3  &    -     &  77.9      &   44.5&14  \\
  4674.60 &   &   62.1   &  0.18   &   -1.06   &  -     &     -    &  85.1      &    -  &14   \\
  4726.026&   &   62.1   &  0.33   &   -1.84   & 56.4   &   -      &  33.2      &   40.8&14  \\
  4791.58:&   &   62.1   &  0.10   &   -1.84   &    -   &   7.14   &  29.9      &   22.0 &14 \\
  6437.64 & Eu II &   63.1   &  1.31   &   -0.27   &   -    &     -    &            &   22.2 & 21  \\
$^\#$  6645.130 &&   63.1    & 1.38 &  +0.20 &  hfs-syn   &   --    &  hfs-syn      &hfs-syn  &21\\
  5007.234: &Er I  &   68.0   &  0.62   &   -0.32   &  156.0 &      -   &    -       &  101.0 &14    \\
  4759.671:&Er II   &   68.1   &  0.00   &   -1.11   & 44.9   &  -       &   30.7     &       - &9   \\
  4820.354:&   &   68.1   &  1.40   &   -0.48   & 107.6  &  73.9    &   102.5    &   68.0 &14  \\
  4757.542:&W I   &   74.0   &  0.36   &   -2.43   &75.2    &  -   &   36.7     &   - & 22    \\
  4048.054:&Os I   &   76.0   &  1.90   &   -1.00   & 47.6   &    -     &   82.0     &   102.3 &12 \\
  4793.996: &  &   76.0   &  0.51   &   -1.99   &   -    &    34.8  &   18.4     &    -&12   \\
  4865.607: &  &   76.0   &  3.79   &    0.31   &  -     &    70.4  &  65.3      &    - &12   \\
4057.807   & Pb I &   82.0   &  1.32   &  -1.71    &  syn   &   --     &  syn      & syn  &4\\
\hline
\hline 
\end{tabular}

1. Wiese et al. (1966), 2. Aoki et al. (2005), 3. Luck, Private communication, 4. Martin et al.(1988), 5. Fuhr et al. (1988), 6. Corliss and Bozman (1962), 7. Hannaford et al. (1982), 8. Biemont et al. (1981), 9. Wolnik and Berthel (1973),
10. Salih and Lawler (1985), 11. Andersen et al. (1975), 12. Corliss and Bozman (1962) adjusted, 13. Arnesen et al. (1977), 14. McEachran and Cohen (1971), 15. Meggers et al. (1975), 16.  Lage and Whaling (1976), 17. Ward et al (1985), 18.  Ward et al. (1985) (WVA), 
19. Hannaford et al. (1982), 20. Saffman and Whaling (1979), 21. Biemont et al. (1982), 22. Obbarius and Kock  (1982).
\end{table*}
}

\end{document}